\documentclass[12pt,letterpaper]{article}
\pdfoutput=1

\usepackage{graphicx,array}
\usepackage{url}
\usepackage{color}
\usepackage{latexsym}
\usepackage{amsthm}
\usepackage{amsmath}
\usepackage{amssymb}
\usepackage{amsfonts}
\usepackage[numbers,sort&compress]{natbib}
\usepackage{bm}
\usepackage{slashed}
\usepackage{mathrsfs}
\usepackage{enumerate}
\usepackage{tikz}
\usepackage{siunitx}
\usepackage{mdframed}
\usepackage{setspace}  
\usepackage{esvect}
\usepackage{esint}
\usepackage{subcaption}
\usepackage[normalem]{ulem}
\usepackage{setspace}

\usepackage[utf8x]{inputenc}%
\usepackage{tcolorbox}%

\usepackage{hyperref} %Automatically links \label and \ref commands; Always load last

\hypersetup{
    colorlinks=true,       % false: boxed links; true: colored links
    linkcolor=red,          % color of internal links
    citecolor=blue,        % color of links to bibliography
    filecolor=magenta,      % color of file links
    urlcolor=blue           % color of external links
}
\usepackage[all]{hypcap} %Link navagates to top of figure instead of caption (below fig)

\usepackage{natbib}
\newcommand{\nc}{\newcommand}

\nc{\beq}{\begin{equation}}  
\nc{\eeq}{\end{equation}}  
\nc{\beqa}{\begin{eqnarray}}  
\nc{\eeqa}{\end{eqnarray}}  
\nc{\bit}{\begin{itemize}}  
\nc{\eit}{\end{itemize}}  

\newcommand{\eg}{{\it e.g.}}

\usepackage{floatrow}
\newfloatcommand{capbtabbox}{table}[][\FBwidth]

\usepackage{blindtext}

\newcommand{\ds}{\displaystyle}

\title{ 
{\bf Global Electroweak Symmetric Vacuum}
\author{\large Yang Bai$^{\,\star}$, Seung J. Lee$\,^\diamond$, Minho Son$^{\,\dagger}$, and Fang Ye$^{\,\dagger}$}
\date{\small \it 
$^\star$Department of Physics, University of Wisconsin-Madison, Madison, WI 53706, USA\\
$^\diamond$Department of Physics, Korea University, Seoul 136-713, Korea \\
$^\dagger$Department of Physics, Korea Advanced Institute of Science and Technology, \\
291 Daehak-ro, Yuseong-gu, Daejeon 34141, Republic of Korea\\
}
}

\begin{document}

\maketitle

\setlength{\parskip}{0.2ex}

\begin{abstract}	
Although the Higgs potential in the Standard Model (SM) contains only a simple electroweak symmetry breaking vacuum in the small field region, additional metastable or global vacua could exist in models beyond the SM. In this paper, we study one intriguing scenario with an additional electroweak symmetric vacuum that could be the global one.  For the thermal  universe ending at the current metastable vacuum, the electroweak symmetry should stay non-restored at high temperatures. We realize the scenario in a model with Higgs-portal couplings to SM singlet scalars with approximately global $O(N)$ symmetries with a large $N$. For a large portion of model parameter space, both the quantum and thermal tunneling rates are suppressed such that our current metastable vacuum is long-lived enough. Our scenario predicts order-one changes for the Higgs self-couplings and a large contribution to the signal of the off-shell Higgs invisible decay. It can be partly probed at the LHC Run 3 and  well tested at the high luminosity LHC.  We also discuss the subcritical (anti-de Sitter) bubbles from the thermal tunneling that could have a large population and interesting cosmological implications.
\end{abstract}

\thispagestyle{empty}  
\newpage  
  
\setcounter{page}{1}

\begingroup
\hypersetup{linkcolor=black,linktocpage}
\tableofcontents
\endgroup
\newpage

%\emailAdd{yangbai@physics.wisc.edu}
%\emailAdd{s.jj.lee@gmail.com}
%\emailAdd{minho.son@kaist.ac.edu}
%\emailAdd{shivaandalice@gmail.com}

%\keywords{}
%\arxivnumber{2001.xxxx}
%\date{\today}

%\newpage
%%%%%%%%%%%%%%%%%%%%%%%%%%%%%%%%%%%%%%%%
%%%%%%%%%%    Main Text starts here
%%%%%%%%%%%%%%%%%%%%%%%%%%%%%%%%%%%%%%%%

%%%%%%%%%%%%%%%%%%%%%%%%%
% INTRODUCTION
%%%%%%%%%%%%%%%%%%%%%%%%%
\section{Introduction}
\label{sec:intro}

\begin{figure}[tph]
\begin{center}
\includegraphics[width=0.5\textwidth]{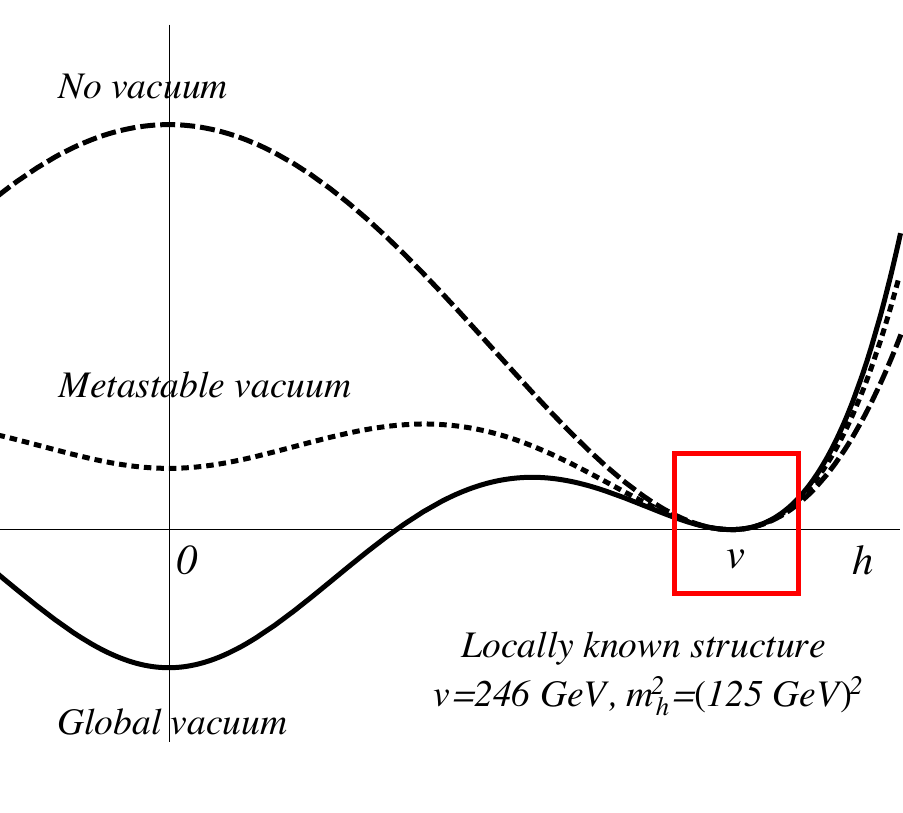}
\caption{\small Representative shapes of the Higgs potential at zero temperature, that share the same locally known structure around $v=246$~GeV (denoted by a box around the Higgs VEV) and have different shapes at the origin.
}
\label{fig:outline}
\end{center}
\end{figure}

With the discovery of the Higgs boson at the LHC~\cite{ATLAS:2012ulh, CMS:2012rpq}, one of the imperative tasks remaining is to understand the nature of the electroweak (EW) symmetry breaking and various aspects of it. Whether the EW symmetry breaking vacuum that our current universe is sitting on is the global vacuum or a metastable one is a fundamental question yet to be answered. In fact, in view of landscape with an exponentially large number of metastable vacua~\cite{Bousso:2000xa, Kachru:2003aw, Susskind:2003kw, Douglas:2003um} in string theory, it will not be too surprising that the EW symmetry breaking vacuum is a metastable one.  However, whether it is a metastable or global one is relatively less explored both theoretically and experimentally. With our current knowledge of the Higgs potential being limited around the local EW symmetry breaking vacuum, the global shape of the Higgs potential is largely unknown and can be qualitatively different from that of the Standard Model (SM), as illustrated in Fig.~\ref{fig:outline}. Our ignorance of the global structure of the Higgs potential also means that we do not have a clear picture of the cosmological evolution of the EW vacuum/vacua and the electroweak phase transition (EWPT).
The possibility of the EW symmetry breaking vacuum being metastable could give rise to a qualitatively different story for the cosmological evolution of the SM fermion and gauge boson masses, which deserves dedicated studies. 
From the theoretical side, this possibility of our current EW symmetry breaking vacuum being a metastable one has been studied mainly in the context with a global vacuum located at a large Higgs field value. This global vacuum in the SM is mainly due to the top quark Yukawa coupling contribution to the renormalization group running of the Higgs quartic coupling~\cite{Sher:1988mj,Degrassi:2012ry}. 
On the other hand, in models beyond the SM, there are many examples with a metastable EW symmetry breaking vacuum.  Some models could have an additional vacuum in the small Higgs field region including the electroweak symmetric (EWS) vacuum at the origin in the Higgs field direction.~\footnote{Note that, even with a positive Higgs mass at the origin, the QCD interaction induced fermion condensation can still break the EW symmetry, although with a much smaller scale~\cite{Quigg:2009xr}. We will neglect the QCD effects for the Higgs potential in this study.} In this paper, we introduce a simple and representative model to realize this scenario. 
For a phenomenologically viable set-up of such scenario, three conditions have been satisfied: a) the lifetime of the metastable EW symmetry breaking vacuum due to quantum tunneling should be at least comparable to the age of the universe; b) our thermal universe should have a symmetry non-restoring (SNR)~\cite{Weinberg:1974hy} evolution at high temperatures such that it will end up at the current metastable vacuum; c) the thermal tunneling rate from the metastable to the global vacuum should be suppressed enough such that the phase transition is uncompleted. 
%%%%%%%%%%%%%%%%%%%%%%%
%%%%%%%%%%%%%%%%%%%%%%%

The early universe evolution for the scenario with a global symmetric vacuum is different from the ordinary EWPT. 
The most common assumption for the ordinary evolution of the EW symmetry is that the EW unbroken phase in the early universe goes through the EWPT to the broken phase as the temperature cools down. 
A lot of attention has been given to this ordinary EWPT story, because the EWPT could be a first-order one to realize the EW baryogenesis~\cite{Cohen:1993nk,Cline:2006ts}. 
Recently, there has been also increased interests in exploring an alternative cosmological history of the universe~\cite{Meade:2018saz,Baldes:2018nel,Glioti:2018roy,Matsedonskyi:2020mlz,Cao:2021yau}, by reviving the old idea of SNR, which is also known as inverse symmetry breaking, where the EWPT doesn't occur at the temperature of the EW scale, but rather at a much higher temperature. There is an SNR in between them, such that the broken vacuum remains to be the one that the universe sits on even for the temperature much above the EW scale.~\footnote{SNR has been adopted to solve other problems such as the monopole ~\cite{Langacker:1980kd,Salomonson:1984rh,Dvali:1995cj} and domain wall~\cite{Dvali:1995cc} problems in Grand Unified Theories.}
A salient feature of a simple model which realizes the EW SNR idea is to introduce SM singlet scalar fields with a large multiplicity of $N$ and a negative portal coupling to the SM Higgs such that one can achieve the goal in an effective field theory with perturbativity under control~\cite{Meade:2018saz,Glioti:2018roy,Chai:2020zgq}.
One attractive feature for the EWPT occurring at a higher temperature is to have an easier implementation of the baryogenesis with relaxed constraints from flavor and CP violation bounds~\cite{Mohapatra:1979qt, Mohapatra:1979vr, Baldes:2018nel,Glioti:2018roy,Matsedonskyi:2020mlz}. 
For our scenarios in this paper, we also have EW SNR at high temperatures and could achieve baryogenesis based on EWPT at higher temperatures. On the other hand, the scenario considered in this paper has the EW breaking vacuum to be the metastable one and the EW unbroken vacuum to be the global one, which will have a different consideration for early universe and also different phenomenological consequences. Interestingly, the SNR is the necessary feature of the scenario considered in this paper, unlike other works on SNR, where SNR is an optional possibility.

In this work, we will consider a model with the minimal extension of the SM with two different copies of scalar singlets that couple to the SM Higgs via a quartic portal coupling. The EW symmetry breaking vacuum with a vacuum expectation value (VEV) $v=246$~GeV is a metastable one, with a global symmetry-preserving vacuum at the origin (without additional vacuum at a large Higgs field value, at least within the cutoff of our effective field theory). This can be regarded as a simple toy model that could be embedded into other models addressing the EW hierarchy problem and expanded to models with multiple fields and a large number of vacua. To be under perturbative control, we will impose approximately $O(N)$ symmetries for the new SM gauge singlet scalars. With a large $N$, we will demonstrate that the perturbativity is kept for the parameter space to achieve the global EW symmetric vacuum. 

For the two types of singlet scalars with separate $O(N)$ symmetries, we have one type responsible for developing a global EW symmetric vacuum from its contribution to the Coleman-Weinberg (CW) potential of the Higgs field, while the other type mainly responsible for SNR in the thermal universe. Given the existence of additional scalars, one could also have non-trivial vacua along their field directions. For instance, in some EWPT models the so-called ``two-step" phase transition has been considered with the final transition from the vacuum with zero Higgs VEV $\langle h \rangle = 0$ and non-zero singlet VEV to the vacuum with $\langle h \rangle = v$ and zero singlet VEV. For this case, one could also study the consequence if the first vacuum with $\langle h \rangle = 0$ is the global vacuum. In this paper, we will focus on the scenario  with the Higgs field as the only order parameter to define different vacua. The advantage of studying scenario is that, besides the merit of simplicity, the Higgs boson properties are modified more dramatically and could be tested more concretely at the LHC in the near future.

From the collider perspective, searching for a deviation of the Higgs trilinear coupling and $Z$-Higgs interactions from the SM predictions can help on reconstructing the Higgs potential, probe the cosmological history of the EW symmetry breaking, and provide a clue for the EW baryogenesis. It serves as one of the important motivations for the future colliders (\eg, 100 TeV proton-proton circular collider). What distinguishes our scenario from other models in the literature is that our model predicts large enough corrections to the Higgs cubic (and quartic) self-couplings, such that it will be testable in the near future both at the LHC Run 3 with around 300\,fb$^{-1}$ luminosity and at the high-luminosity (HL) LHC with 3\,ab$^{-1}$. As the Higgs boson has direct couplings to two copies of scalar fields under separate $O(N)$ symmetries, it turns out that the signal with jets plus missing transverse momentum [via off-shell Higgs invisible decay from vector-boson-fusion (VBF) productions] will be another ``golden channel" to test our scenario at the HL-LHC experiment, or even at the LHC Run 3 if the singlet scalars are relatively light.

For the (uncompleted) transition from the EW symmetry breaking vacuum to the global unbroken vacuum, a first-order phase transition is generic in the model parameter space. We will demonstrate that the thermal tunneling rate is suppressed such that this first-order phase transition is an uncompleted one, which is different from the ordinary EWPT where the phase transition is quickly completed below and around the critical temperature. For our scenario, some nucleated bubbles could have a negative energy density inside the bubble and hence behave as an anti de-Sitter (AdS) bubbles. Those bubbles, mainly subcritical, will shrink and disappear if there is no additional pressure to be against the tension force. For the supercritical bubbles, their run-away speed is generically large~\cite{Bodeker:2017cim,Hoeche:2020rsg,Vanvlasselaer:2020niz} and they could easily occupy a big fraction of the current Hubble volume and overclose the universe if they are produced with a large rate. Therefore, we will estimate their formation rate and use it to constrain the allowed model parameter space.  

Our paper is organized as follows. In Section~\ref{sec:model}, we introduce a model with Higgs-portal couplings to SM singlet scalars with approximately global $O(N)$ symmetries with a large $N$. We list constraints on the couplings to ensure the perturbativity. In Section~\ref{sec:globalvacua}, we establish the global EW symmetric vacuum, while lifting the EW symmetry breaking vacuum to the metastable one, by modifying Higgs potential due to the radiative corrections from new SM singlet scalars. In Section~\ref{sec:SNR}, we study the EW symmetry non-restoring thermal history of the universe to have it end at the current metastable symmetry breaking vacuum. We present two benchmark scenarios (along with the parameter scans around them) which have the metastable vacuum to be long-lived enough against both the quantum and thermal tunneling processes. In Section~\ref{sec:bubble}, we discuss the formation of bubbles including the AdS ones that could be populated during the thermal evolution of the universe and their dynamics. In Section~\ref{sec:pheno}, we discuss how to test our scenario at the LHC Run 3 and HL-LHC. In Appendix~\ref{app:sec:frac:bubbles}, we derive the lower bound on the bounce action of finite temperature to ensure vanishing fraction of the space occupied by supercritical bubbles in the current universe. In Appendix~\ref{app:sec:interbubble:dist}, we estimate the typical distance between subcritical bubbles. In Appendix~\ref{app:sec:RGE}, we solve the renormalization group equations of scalar quartic couplings and derive the Landau pole scale. In Appendix~\ref{app:sec:atlas:recasting}, we recast the ATLAS analysis on the invisible decay of the off-shell Higgs in the VBF channel to derive the current (and future) sensitivity on the parameters of our scenario. 

%%%%%%%%%%%%%%%%%%%%%%%%%
% MODEL
%%%%%%%%%%%%%%%%%%%%%%%%%
\section{Theoretical framework}
\label{sec:model}
The scenario considered in this paper is the extension of the SM by including two copies of SM gauge-singlet real scalar fields, $S$ and $\Phi$, transforming as fundamental representations under $O(N_s)$ and $O(N_\phi)$, respectively, and coupling to the SM via the Higgs-portal couplings. The scalar potential is
\begin{equation}
\label{eq:lag1}
\begin{split}
  V_{\rm tree} =&\ - \mu^2_h\,H^\dagger H + \lambda_h\,(H^\dagger H)^2
  \\[1.5pt]
&+  \lambda_{hs}\, H^\dagger H\, S^2+ \frac{m^2_s }{2} \, S^2 + \frac{\lambda_s}{4}\,(S^2)^2
+ \lambda_{h\phi}\, H^\dagger H \,\Phi^2  + \frac{m^2_\phi}{2} \,\Phi^2 + \frac{\lambda_\phi}{4} \, (\Phi^2)^2~,
\end{split}
\end{equation}
where $S^2 \equiv S^{\rm T} S$ and $\Phi^2 \equiv \Phi^{\rm T} \Phi$. The self-interacting quartic coupling of the SM Higgs field has a numeric value of $\lambda_h = m_h^2/(2v^2)\approx 0.13$ with $m_h\approx 125$~GeV and $v=246$~GeV. Here, we have ignored the quartic coupling between  the two singlet fields $\lambda_{s\phi} S^2 \Phi^2$ (this interaction can be generated at one-loop level with $\lambda_{s\phi} \sim \lambda_{hs} \lambda_{h\phi} / (16\pi^2)$, which is small enough to affect our later results).
In terms of the neutral component of the Higgs field $H=(0, h)^{\rm T}/\sqrt{2}$, the potential takes the form,
\begin{equation}
\begin{split}
  V_{\rm tree} =&\ - \frac{\mu^2_h}{2} \,h^2 + \frac{\lambda_h}{4} \, h^4 
  \\[1.5pt]
  & +  \frac{\lambda_{hs}}{2}\,h^2 S^2 + \frac{m^2_s }{2} \, S^2 + \frac{\lambda_s}{4} \,(S^2)^2
  +  \frac{\lambda_{h\phi}}{2}\, h^2\,\Phi^2  + \frac{m^2_\phi}{2}\, \Phi^2 + \frac{\lambda_\phi}{4}\, (\Phi^2)^2~.
\end{split}
\end{equation}
In our study, we will only consider the parameter space with the SM Higgs developing a non-zero VEV $\langle h \rangle = v = 246$~GeV for the vacuum of our current universe sitting on and zero VEVs for both $S$ and $\Phi$. At tree level, this requires $\mu_h^2 > 0$, $m_s^2 + \lambda_{hs} v^2 > 0$ and $m_\phi^2 + \lambda_{h \phi} v^2 > 0$.~\footnote{We have also checked the multi-field potential at one-loop level to ensure that there is no vacuum with nonzero $S$ and $\Phi$ VEVs.} In order to prevent the EW symmetry from being restored at high temperatures, we will take $\lambda_{hs} < 0$ and positive values for other quartic couplings: $\lambda_h>0$, $\lambda_s >0$, $\lambda_\phi >0$, $\lambda_{h\phi} >0$. 
Requiring the potential is bounded from below with $\lambda_{hs} < 0$, additional constraints on the quartic couplings can be obtained by focusing on the large field region
\begin{equation}
\begin{split}
  V_{\rm tree} \approx&\ \frac{\lambda_h}{4} \, h^4 + \frac{\lambda_{hs}}{2}\, h^2 S^2 + \frac{\lambda_s}{4}\, (S^2)^2~
 \\[1.5pt]
  =&\ \frac{1}{4} \left ( \sqrt{\lambda_h} \, h^2  - \sqrt{\lambda_s} \, S^2 \right )^2 + \frac{1}{2} \, h^2 S^2 \left ( \lambda_{hs} + \sqrt{\lambda_h} \sqrt{\lambda_s} \right )~,
\end{split}
\end{equation}
and it leads to the condition,
\begin{equation}
\label{eq:stability:hs-h-s}
- \sqrt{\lambda_h} \sqrt{\lambda_s} \leq  \lambda_{hs} < 0   ~.
\end{equation}
Demanding the perturbativity puts a series of constraints on the quartic couplings. The naive dimensional analysis (NDA) using the diagrammatic estimate gives rise to the perturbativity bound on the portal couplings, $\lambda_{hs}$ and $\lambda_{h\phi}$, respectively,
\begin{equation}
\label{eq:perturbative-lambda-hs-hphi}
 \frac{\sqrt{N_s}\, |\lambda_{hs}|}{16\pi^2}~, \quad \frac{\sqrt{N_\phi} \,\lambda_{h\phi}}{16\pi^2} < \mathcal{O}(1)~.
\end{equation}
Similarly the perturbativity on self quartic couplings, $\lambda_s$ and $\lambda_\phi$, gives rise to
\begin{equation}
\label{eq:perturbative-diag-lambda}
\begin{split}
 \frac{N_s \,\lambda_s}{16\pi^2}~,\quad \frac{N_\phi \,\lambda_\phi}{16\pi^2} < \mathcal{O}(1)~.
\end{split} 
\end{equation}
As long as the perturbativity of the quartic couplings in Eqs.~(\ref{eq:perturbative-lambda-hs-hphi}) and~(\ref{eq:perturbative-diag-lambda}) and the Higgs quartic coupling is kept up to a very high energy scale, the stability condition in Eq.~(\ref{eq:stability:hs-h-s}) will also be easily satisfied for running couplings. The lowest energy scale where any of the perturbativity bounds breaks down can be treated as the ultraviolet cutoff scale $\Lambda_{\rm UV}$ of our effective field theory model. After examining the renormalization group equations of the quartic couplings (see Appendix~\ref{app:sec:RGE} for more details), we find that either $\lambda_{hs}$ or $\lambda_{h\phi}$ can develop a Laudau pole above the EW scale. For instance, assuming a very large multiplicity, $N_\phi \gg 1$, and a negligible initial value of $\lambda_\phi$ at the EW scale, one can derive the Landau pole scale associated with the coupling $\lambda_{h\phi}$ as
\begin{equation}
\label{eq:landau-scale}
  \Lambda_{\rm L} \sim v \exp \left [ \frac{16\pi^2}{\sqrt{8(2N_\phi+4)} } \frac{1}{\lambda_{h\phi}(v)} \right ]~.
\end{equation} 
Similar argument applies to the couplings related to $S$. As will be discussed later in detail, $N_\phi \lambda^2_{h\phi}(v)$ is constrained to satisfy $N_\phi \lambda^2_{h\phi}(v) > 41$ to exhibit the global electroweak symmetric vacuum at the origin, and thus the cutoff scale of our effective field theory is estimated to be $\Lambda_{\rm UV} \sim \Lambda_{\rm L} \lesssim 120$ TeV.

%%%%%%%%%%%%%%%%%%%%%%%%%
% Global electroweak vacua 
%%%%%%%%%%%%%%%%%%%%%%%%%
\section{Global electroweak symmetric vacuum}
\label{sec:globalvacua}
Using the on-shell renormalization scheme in Landau gauge, the one-loop Coleman-Weinberg potential of the Higgs field from the fields $S$ and $\Phi$ is given by
\begin{equation}\label{eq:Veff:toy}
  V_{\rm CW} = \sum_{X=S,\, \Phi} \frac{N_X}{64\pi^2} \left [ m^4_X(h) \left ( \ln \frac{m^2_X(h)}{m^2_X(v)} - \frac{3}{2} \right ) + 2 m^2_X (h)\, m^2_X(v)  \right ]~, 
\end{equation}
where the field dependent masses are $m^2_X(h) = m^2_X + \lambda_{hX} h^2$. The total effective potential is $V_{\rm eff} \equiv V_{\rm tree} + V_{\rm CW}$. We will consider parameter space to have the electroweak symmetry breaking vacuum $\langle h \rangle = v$ to be a metastable one. The global vacuum of the total potential $V_{\rm eff}$ is located at $\langle h \rangle = 0$ with unbroken EW symmetry. At zero temperature, the transition time from our current symmetry breaking vacuum to the global EW-symmetric vacuum is required to be longer than the age of the universe. 
Since we have adopted on-shell renormalization scheme for the CW potential, the physical Higgs mass and VEV is determined by the tree level potential. Without the CW potential, the tree-level potential only admits one ordinary symmetry breaking vacuum. The situation can be different if the CW potential becomes significant. 

Two non-trivial conditions to have a metastable EW symmetry breaking vacuum with a global vacuum at the origin are
\begin{equation}\label{eq:metavac:cons:all}
m^2_h (0) = \left . \frac{d^2V_{\rm eff}}{dh^2} \right |_{h=0} > 0 ~, \quad 
\epsilon(v) \equiv V_{\rm eff} (v) - V_{\rm eff}(0) > 0~.
\end{equation}
For a better analytic understanding, we rewrite the first condition as
\begin{equation}\label{eq:metavac:cons1}
\begin{split}
m^2_h (0) =  \left . \frac{d^2V_{\rm eff}}{dh^2} \right |_{h=0} =& - \lambda_h v^2 + \frac{N_s \, \lambda_{hs}}{16\pi^2}\,m_s^2 \left [ \ln \left ( \frac{1}{1 + (\lambda_{hs} v^2)/m_s^2} \right ) + \frac{\lambda_{hs} \, v^2} {m_s^2} \right ] 
  \\[2.5pt]
 &\hspace{1.0cm} + \frac{N_\phi\, \lambda^2_{h\phi}}{16\pi^2}\, v^2 \left [ \frac{m_\phi^2}{\lambda_{h\phi}\, v^2} \ln \left ( \frac{m_\phi^2/(\lambda_{h\phi} v^2)}{m_\phi^2/(\lambda_{h\phi} v^2) + 1 } \right ) + 1 \right ]>0~,
\end{split}
\end{equation}
where $\mu^2_h = \lambda_h v^2$  was used. Terms inside the bracket for $\phi$ is roughly order one positive value as long as $m_\phi^2/(\lambda_{h\phi} v^2) < \mathcal{O}(10^{-1})$, and the second term for $s$ is negative definite.
In our benchmark scenarios we will focus on the parameter space satisfying the following properties, 
\begin{equation}\label{eq:benchmark:assumptions}
m_s^2 \sim v^2~,\quad |\lambda_{hs}| \ll 1~,\quad \alpha \equiv {m_\phi^2 \over \lambda_{h\phi} v^2} \ll 1~.
\end{equation}
When properties in Eq.~(\ref{eq:benchmark:assumptions}) are assumed, the inequality in Eq.~(\ref{eq:metavac:cons1}) becomes approximately
\begin{equation}\label{eq:metavac:cons1:approx}
 m^2_h(0) \approx - \lambda_h v^2 + \frac{N_s\, \lambda^3_{hs}}{32\pi^2}  \frac{v^4}{m_s^2} + \frac{N_\phi \,\lambda^2_{h\phi}}{16\pi^2}v^2
\approx - \lambda_h v^2 + \frac{N_\phi\, \lambda^2_{h\phi}}{16\pi^2} v^2 > 0~,
\end{equation}
where we have assumed $\sqrt{N_s} |\lambda_{hs}|^{3/2}/\sqrt{2} \ll \sqrt{N_\phi} \lambda_{h\phi}$ (with $m_s^2 \sim v^2$) based on the previous $|\lambda_{hs}| \ll 1$ assumption, which can make the positivity of $m^2_h(0)$ better satisfied as it suppresses the negative contribution from $S$.
The condition in Eq.~(\ref{eq:metavac:cons1:approx}) provides a lower bound on the combination of quartic $\lambda_{h\phi}$ and $N_\phi$, namely $N_\phi \lambda^2_{h\phi}$, 
\begin{equation} \label{eq:constraint:pos:mhsq}
 N_\phi\, \lambda_{h\phi}^2 > 16 \pi^2\, \lambda_h \simeq 21~.
\end{equation}
Similarly the second condition in Eq.~(\ref{eq:metavac:cons:all}) can be rewritten as
\begin{equation}\label{eq:metavac:cons2}
\begin{split}
 \epsilon(v) =& - \frac{1}{4}\lambda_h\,v^4 + \frac{N_s}{128\pi^2} \,m_s^4
 \left [ \frac{\lambda_{hs}^2\,v^4}{m^4_s} - 2 \frac{\lambda_{hs} \,v^2}{m^2_s} - 2 \ln \left ( \frac{1}{1 + (\lambda_{hs} \,v^2)/m_s^2} \right )  \right ]
\\[2.5pt]
& \hspace{1.6cm} + \frac{N_\phi\, \lambda_{h\phi}^2}{128\pi^2} \, v^4
\left [  1 - 2 \frac{m^2_\phi}{\lambda_{h\phi}\, v^2} - 2 \frac{m^4_\phi}{\lambda_{h\phi}^2 \,v^4} \ln \left ( \frac{m_\phi^2/(\lambda_{h\phi} \,v^2)}{m_\phi^2/(\lambda_{h\phi} \,v^2) + 1 } \right )  \right ] > 0~,
\end{split}
\end{equation}
where the bracket for $\phi$ is roughly order-one positive value as long as $\alpha = m_\phi^2/(\lambda_{h\phi} v^2) < \mathcal{O}(10^{-1})$ and the term for $s$ is negative definite. When Eq.~(\ref{eq:benchmark:assumptions}) is assumed as above, the above inequality becomes approximately
\begin{equation}\label{eq:metavac:cons2:approx}
\epsilon(v) 
\approx  - \frac{1}{4}\lambda_h v^4 +  \frac{m_s^4 N_s}{192\pi^2} \left ( \frac{\lambda_{hs} v^2}{m_s^2} \right )^3 + \frac{N_\phi \, \lambda_{h\phi}^2}{128\pi^2} \, v^4
\approx  - \frac{1}{4}\lambda_h v^4 + \frac{N_\phi \, \lambda_{h\phi}^2}{128\pi^2} \, v^4 > 0~,
\end{equation}
with the assumption of $\sqrt{N_s} |\lambda_{hs}|^{3/2}/\sqrt{3/2} \ll \sqrt{N_\phi} \lambda_{h\phi}$ (with $m_s^2 \sim v^2$) to suppress negative contribution from $S$.~\footnote{
The negative sub-leading contribution from $S$ reduces the size of the positive $\epsilon(v)$ [also the value of $m_h^2(0)$] for lower $m_s^2$ ($m^2_s \lesssim v^2$). Accordingly, the critical temperature at which two vacua become degenerate will be raised for a higher value of $m_s^2$.
%the transition rate will the maximum at the temperature closer to the critical $T$ than the case with a smaller contribution from $S$.
} Again the positivity of $\epsilon(v)$ in Eq.~(\ref{eq:metavac:cons2:approx}) depends only on $N_\phi \lambda^2_{h\phi}$ and leads to a lower bound on it,
\begin{equation} \label{eq:constraint:pos:epsilon}
\mbox{\it global vacuum at }\langle h \rangle = 0:\quad N_\phi\,\lambda_{h\phi}^2 > 32\pi^2\, \lambda_h \simeq 41~,
\end{equation}
which can easily satisfy the perturbativity constraints in \eqref{eq:perturbative-lambda-hs-hphi}. 
Noting that both conditions in Eq.~(\ref{eq:metavac:cons:all}) under our assumption in Eq.~(\ref{eq:benchmark:assumptions}) place a lower bound on $N_\phi \lambda_{h\phi}^2$. Interestingly, since the bound in Eq.~(\ref{eq:constraint:pos:epsilon}) is more stringent than the one from Eq.~(\ref{eq:metavac:cons1:approx}), it allows the inverted relation between the two vacua, namely the symmetry unbroken vacuum at the origin is metastable and the symmetry breaking vacuum is the global one, for the following range of $N_\phi\, \lambda_{h\phi}^2$,
\begin{equation} \label{eq:vac:inverted}
\mbox{\it metastable vacuum at }\langle h \rangle = 0:\quad  41 \simeq 32 \pi^2\, \lambda_h > N_\phi\, \lambda_{h\phi}^2 > 16 \pi^2\, \lambda_h \simeq 21~.
\end{equation}
As we will discuss below, the experimental probe for the existence of the metastable unbroken vacuum at the origin in our scenarios will be challenging even at the HL-LHC.

%%%%%%%%%%%%%%%%%%%%%%%%%
% Electroweak symmetry nonrestoratioin
%%%%%%%%%%%%%%%%%%%%%%%%%
\section{Symmetry non-restoration at high temperatures}
\label{sec:SNR}
After demonstrating the existence of parameter space to have $\langle h \rangle = 0$ as the global vacuum, one also needs to worry about why our current universe ends up at the electroweak symmetry breaking vacuum with $\langle h \rangle = v$. In this section, we want to demonstrate that with the choice of $\lambda_{hs} < 0$ and a related condition, the electroweak symmetry could stay at the broken one all the way up to a high temperature. This symmetry non-restoration has been pointed out a while ago by Weinberg~\cite{Weinberg:1974hy} and explored in Refs.~\cite{Mohapatra:1979qt,Mohapatra:1979vr,Dvali:1995cj} and more recently in Refs.~\cite{Meade:2018saz,Baldes:2018nel,Glioti:2018roy,Matsedonskyi:2020mlz,Cao:2021yau}. 

At high temperatures of the early universe, the thermal corrections to the scalar effective potential can change the shape of the potential so that our current metastable vacuum could become a global vacuum.  
We will study the phenomenological implication of this possibility in this work.
In the high $T$ approximation, the thermal corrections are dominated by the scalar thermal masses. 
Neglecting SM interaction contributions, one has~\cite{Dolan:1973qd}
\beqa\label{eq:Vth:toy:highT}
  \Delta V_{\rm thermal} &=& \frac{1}{24} \left (N_s\, \lambda_{hs}+ N_\phi \,\lambda_{h\phi} \right ) T^2 h^2 \nonumber \\
  && \hspace{-1cm} + \, \frac{1}{24} \left[4\,\lambda_{hs}+ (N_s +2) \,\lambda_s \right] T^2 S^2 \, + \,  \frac{1}{24} \left[4\,\lambda_{h\phi}+ (N_\phi +2 ) \,\lambda_\phi \right] T^2 \Phi^2 + \cdots
  ~.
\eeqa
As long as the thermal correction in Eq.~(\ref{eq:Vth:toy:highT}) to the effective potential is the dominant one, the Higgs mass could stay negative around the vacuum $\langle h \rangle  = v$ for a proper choice of $N_s\, \lambda_{hs}+ N_\phi \,\lambda_{h\phi} $. Then, the electroweak symmetry will not be restored up to an arbitrarily high temperature (at least to the cutoff scale of our effective field theory), if satisfying the condition
\begin{equation}\label{eq:constraint:neg:Vth}
 N_s\, \lambda_{hs} + N_\phi \lambda_{h\phi} < 0~,
\end{equation}
where we choose $\lambda_{h s}$ to be negative ($\lambda_{h s} < 0$) to satisfy the above inequality condition. To have no additional saddle points in the $S$ and $\Phi$ directions, we impose one more condition
\beq
\label{eq:constraint:neg:Vth2}
4\,\lambda_{hs}+ (N_s + 2) \,\lambda_s  > 0 ~.
\eeq

Combining Eqs.~(\ref{eq:metavac:cons1:approx}),~(\ref{eq:constraint:pos:epsilon}),~(\ref{eq:constraint:neg:Vth}), and \eqref{eq:constraint:neg:Vth2}, the condition to have a global electroweak symmetric vacuum at both zero and high temperatures is 
\begin{equation}\label{eq:lamhphiwindow:highT}
 \frac{6.41}{\sqrt{N_\phi}}\approx \frac{4\pi \sqrt{2\lambda_h}}{\sqrt{N_\phi}}  < \lambda_{h\phi} < - \frac{N_s}{N_\phi} \lambda_{h s}\,\quad \mbox{and}\quad \lambda_s > - \frac{4\,\lambda_{hs}}{N_s + 2}~.
\end{equation}
The upper bound on $\lambda_{h\phi}$ in Eq.~(\ref{eq:lamhphiwindow:highT}) suggests that asymmetric multiplicities, $N_s > N_\phi$, as $\lambda_{h\phi} > |\lambda_{hs}|$ is preferred based on our assumptions in \eqref{eq:benchmark:assumptions}.
While the lower bound on $\lambda_{h\phi}$ in Eq.~(\ref{eq:lamhphiwindow:highT}) is robust based on our approximation, the upper bound can be relaxed in a situation when the negative thermal correction from the ring term becomes sizable~\cite{Kraemmer:2003gd}. For instance, the ring term from $S$ and $\Phi$ is
\begin{equation}
 V_{\rm ring} = \sum_{x=s,\phi} \frac{N_x T}{12\pi} \left( m_x^{3}(h) - [m_x^2(h) + \Pi_x (0)]^{3/2} \right)~,
\end{equation}
where $m^2_s(h) = m^2_s + \lambda_{hs} h^2$ and the Debye mass $\Pi_s(0) = \frac{1}{12}\left [ 4\lambda_{hs}+ (N_s+2) \lambda_s \right ] T^2$ (similarly for $\Phi$ with the replacement of $s\rightarrow \phi$). When the values of $\lambda_\phi$ and $\lambda_s$ are chosen to be appropriate values so that the ring term contributes sizable to a negative Higgs mass-squared,~\footnote{For the purpose of analytic understanding, we can expand the ring term in either low or high $T$. The leading term in $h$, ignoring unimportant $h$-independent terms, is given by
\begin{equation}
V_{\rm ring} \approx \left\{
\begin{array}{ll}
-  \displaystyle \sum_{x=s,\phi}\frac{N_xT^3}{96\pi} \left\{ \left [ 4 \lambda_{hx} + (2+N_x)\lambda_x \right ] \sqrt{m_x^2 + \lambda_{hx} h^2}\, \right\}  \quad ~ (\text{low-}T),
\vspace{0.4cm}
\\[3.0pt]
- \displaystyle  \sum_{x=s,\phi} \frac{T^2 h^2}{16\sqrt{3}\pi} \Big [ N_x \lambda_{hx} \sqrt{4\lambda_{hx} + (2+N_x)\lambda_x}  \Big ] \quad ~ (\text{high-}T).
 \end{array}
 \right.
\end{equation}
Given that $\lambda_{hs}$ is taken to be negative and the constraint from Eq.~(\ref{eq:constraint:neg:Vth2}), the ring term from $\Phi$ in either case can contribute sizably to the negative Higgs mass-squared. Nevertheless, the negative $\lambda_{hs}$ still helps for the symmetry non-restoration as it reduces the overall size of the Higgs mass-squared term [see Eq.~(\ref{eq:Vth:toy:highT}) for the case in the high-$T$ limit] which competes with the ring term. 
} $\lambda_{h\phi}$ can be pushed to a higher value than $-\lambda_{hs}$ even for the case with $N_s \simeq N_\phi$, while maintaining symmetry non-restoration. In our numerical calculations, we do not work in the high-temperature limit, but include all temperature dependence for the effective potential (see Refs.~\cite{Delaunay:2007wb,Jain:2017sqm}).

Within the parameter space to have our current universe sitting at a metastable EW symmetry breaking vacuum, one needs to make sure that the current universe will not tunnel to the universe with the global EW symmetric vacuum, both at the current era and in the early universe with higher temperatures. 
The tunneling rates per unit volume for both quantum and thermal ones are~\cite{Coleman:1977py,Linde:1980tt}
\begin{equation}
\label{eq:toy:S3vsS4}
\left\{ 
\begin{array} {ll} 
\Gamma_4 \simeq \ds\frac{1}{R^4} \left (\displaystyle\frac{S_4}{2\pi}\right )^2 \exp\left (-S_4 \right )   & \quad T \approx 0~, 
\vspace{0.4cm}\\[5.0pt]
\Gamma_3 \simeq T^4 \left [\displaystyle\frac{S_3(T)}{2\pi T} \right ]^{3/2}  \exp \left[ -\displaystyle{S_3(T)/T} \right]  &\quad T \neq 0~,
\end{array}
\right.
\end{equation}
where $R$ is the size of the $O(4)$ critical bubbles and is related to the parameters in the effective potential. There is a lower bound on $S_4$ to have our current EW symmetry breaking vacuum $\langle h\rangle=v$ not to tunnel into the global vacuum at $\langle h\rangle=0$. Roughly, this requires $\Gamma_4 < H_0^4$ with $H_0$ as the Hubble scale at the current universe. Using $R \sim 1/v$, this condition can be translated into $S_4 > S_4^{\rm min} \simeq \ln[v^4/H_0^4] + 2\ln[\ln(v^4/H_0^4)/(2\pi)] \approx 416$. In our later numerical calculations, we will show that it is fairly easy to satisfy this long-lived-enough condition for our current metastable vacuum. 

At early universe with a higher temperature, the thermal tunneling may have faster transition rates or with $S_3(T)/T < S_4$ for some temperatures. This means that one also needs to satisfy another condition to make sure that the metastable vacuum at $\langle h\rangle = v(T) \neq 0$ does not thermally tunnel to the global one at $\langle h\rangle = 0$. There is a minimum value for $S_3(T)/T$ at the temperature $T_m$. After taking into account  the bubble expansion (see details in Appendix~\ref{app:sec:frac:bubbles}), we derive a lower bound  $S_3(T_m)/T_m \gtrsim  \ln  \left(M_{\rm pl} T^2_m t_0^3 \right) + \frac{3}{2} \ln \left[ \frac{2}{3} \ln (M_{\rm pl} T^2_m t_0^3)\right] \approx 339$ for $M_{\rm pl} = 2.43\times 10^{18}$~GeV, $T_m \simeq 10$~GeV and the age of our universe $t_0\simeq 13.7$~Gyr. Noting that the numerical number 339 is different from the number $\sim 150$ for the ordinary EW phase transition, where the phase transition is completed at the phase transition temperature around the EW scale. For our case, the phase transition is required to be an {\it uncompleted} one, so that the generated bubbles have a longer time to expand and occupy more space. For some strong first-order EW phase transitions, the phase transition occurs shortly after the critical temperature $T_c$.  The energy difference $\epsilon(T)$~\footnote{For the ordinary EW phase transition, $\epsilon (T) =  V_{\rm eff}(v,T)  - V_{\rm eff}(0,T) < 0$ since the EW symmetry breaking vacuum becomes the global one for $T < T_c$. } is smaller than the barrier size,~\footnote{In the discussion of~\cite{Coleman:1977py} which should be qualitatively applicable to our case, this criteria for the validity is equivalent to that the bubble radius should be much larger than the wall thickness, $R \sim 3S_1/\epsilon \gg (V''(0))^{-1/2}$.} and thus the thin-wall approximation can be applied. On the contrary, in our situation, the thin-wall approximation is not necessarily satisfied, and we will rely on the numerical simulation to calculate $S_3(T)/T$ for different temperatures.  More specifically, we will the $\mathsf{Mathematica}$~\cite{Mathematica} package {\sc FindBounce}~\cite{Guada:2020xnz} to calculate both $S_4$ and $S_3(T)/T$.

%%%%%%%%%%%%%%%%%%%%%%%
\begin{figure}[th!]
\begin{center}
\includegraphics[width=0.50\textwidth]{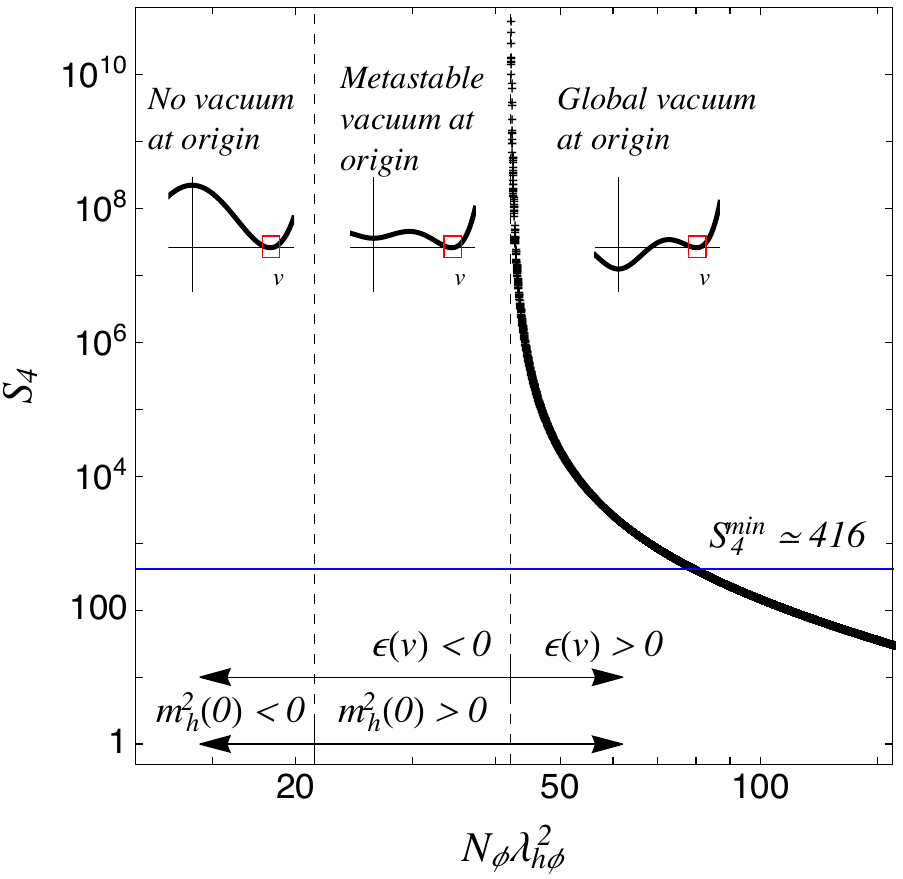}
\caption{\small The bounce action $S_4$ as a function of $N_\phi \lambda^2_{h\phi}$ at zero temperature. Other parameters are fixed to be $N_s = 1500$, $\lambda_{hs}=-0.1$, $\alpha \equiv m_\phi^2 /( \lambda_{h\phi} v^2) = 0.001$.  The peak value is around $N_\phi \lambda^2_{h\phi} \sim (4\pi \sqrt{2\lambda_h})^2\approx 41$ below which $\epsilon(v)$ switches the sign or the global vacuum changes from $\langle h \rangle = 0$ to $\langle h \rangle = v$. For $N_\phi \lambda_{h\phi}^2 > 80$, the current metastable vacuum with $\langle h \rangle = v$ has a short lifetime to tunnel to the global one or $S_4 < S_4^{\rm min}\simeq 416$.}
\label{fig:bounce:B4:NphiLamphi}
\end{center}
\end{figure}
%%%%%%%%%%%%%%%%%%%%%%%

Under the aforementioned assumption on $\lambda_{hs}$, $|\lambda_{hs}| \ll 1$ in Eq.~(\ref{eq:benchmark:assumptions}) and $m_s^2 \sim v^2$, the effective potential at zero temperature is quite insensitive to $N_s$ (its dependence will be important for finite temperature). Two most relevant parameters, $N_\phi$ and $\lambda_{h\phi}$, affect the shape of the zero temperature effective potential (thus the size of $S_4$) through one combination, $N_\phi \lambda^2_{h\phi}$. The calculated $S_4$ as a function of  $N_\phi \lambda^2_{h\phi}$ is shown in Fig.~\ref{fig:bounce:B4:NphiLamphi}. For $N_\phi \lambda^2_{h\phi}$ above or below 41, we have the global vacuum at $\langle h \rangle =0$ or $\langle h \rangle = v$ [see also Eq.~(\ref{eq:vac:inverted})]. When $N_\phi \lambda^2_{h\phi} < 21$, there is no vacuum at $\langle h\rangle = 0$. In the blue horizontal line, we show the minimum value $S_4^{\rm min} \approx 416$, which intersects the black curve at $N_\phi \lambda^2_{h\phi} \approx 80$. When $S_4 >  S_4^{\rm min}$, the current metastable vacuum at $\langle h \rangle = v$ is longer than the age of universe. Therefore, the $41 < N_\phi \lambda^2_{h\phi} < 80$ is the range of parameter space to have a ``safe" global vacuum at origin at zero temperature. 

\begin{figure}[th!]
\begin{center}
\includegraphics[width=0.48\textwidth]{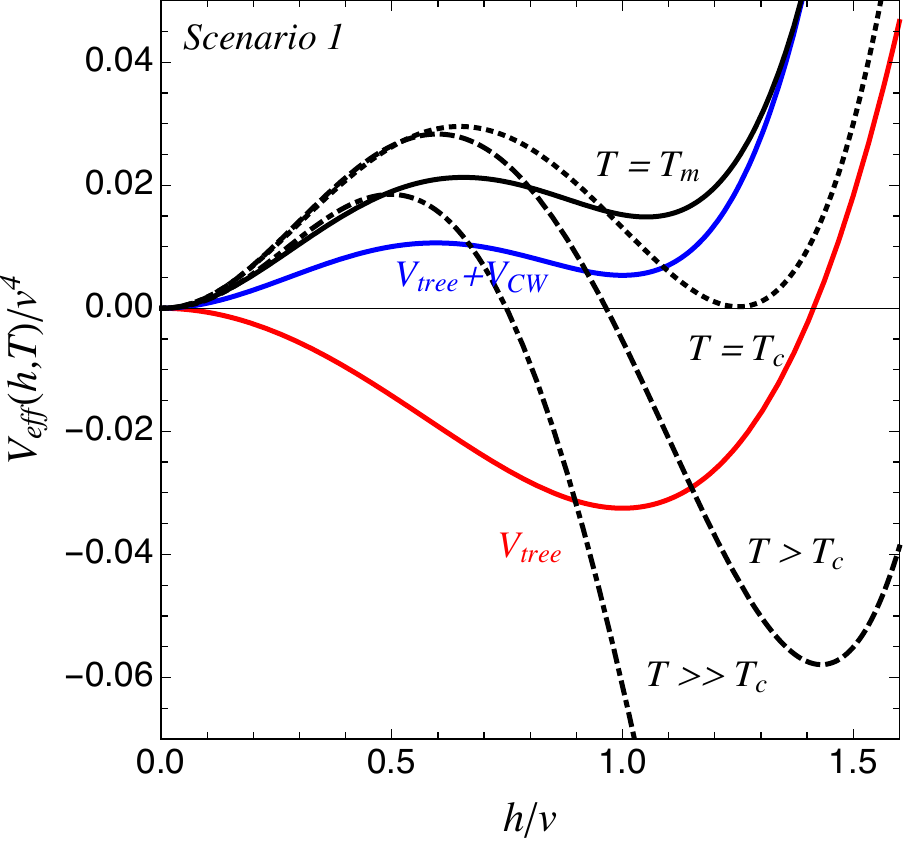} \quad
\includegraphics[width=0.48\textwidth]{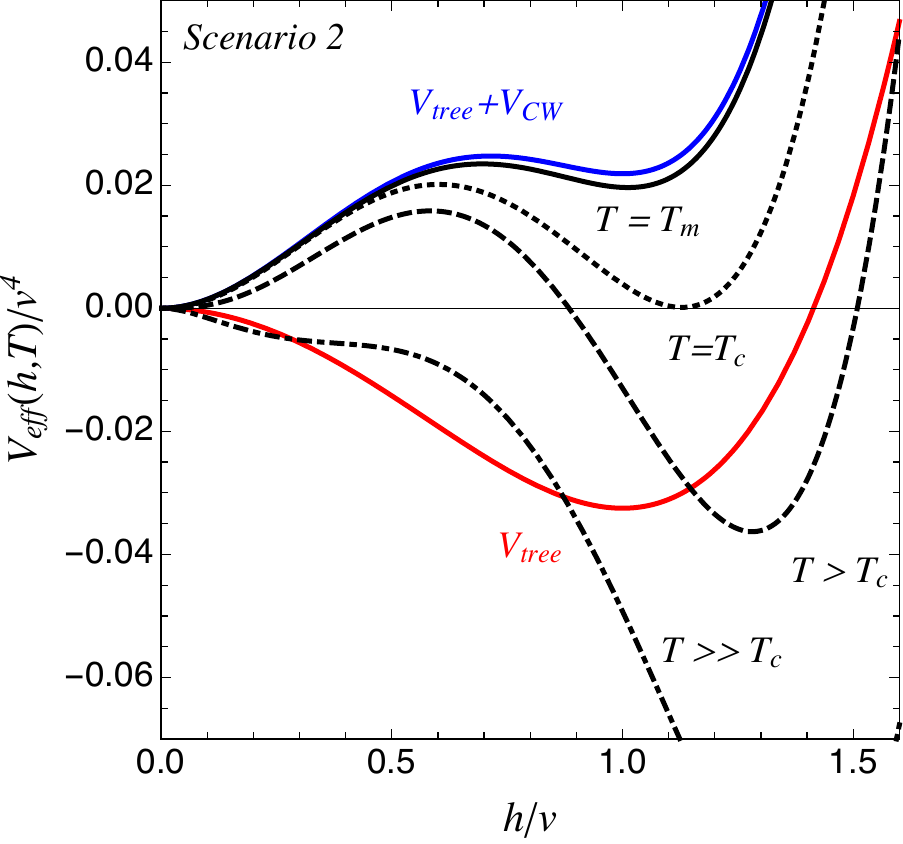}
\caption{\small The effective potential of the benchmark scenario 1 (left) and benchmark scenario 2 (right) in Table~\ref{tab:benchmark}. The pure SM tree-level potential at $T=0$ is shown in solid red line, while the one-loop corrected one at $T=0$ is shown in the solid blue line. The effective potential around $T=T_m < T_c$, where $S_3(T)/T$ takes a minimum value, is shown in solid black line. Similarly, other curves are for $T=T_c$ (dotted black), $T > T_c$ (dashed black), and $T \gg T_c$ (dot-dashed black). 
}
\label{fig:benchmark:Tevolution}
\end{center}
\end{figure}

At high temperatures above a critical temperature $T_c$ when the two vacua are degenerate, the universe prefers sitting at the EW symmetry breaking vacuum. 
As the universe cools down below the critical temperature or $T<T_c$, the symmetry breaking vacuum has a higher free energy than the unbroken one, so that the symmetry breaking vacuum  becomes metastable. As examples, we show two benchmark scenarios with the corresponding parameter values in Table~\ref{tab:benchmark} and their behaviors of the effective potential in Fig.~\ref{fig:benchmark:Tevolution}. The two benchmark points are chosen to have a non-symmetric case ($N_s > N_\phi$) for the benchmark point 1 and a symmetric case ($N_s = N_\phi$) for the benchmark point 2. For the point 1, the ring-term contribution is not important, while for the point 2 with a larger value of $\lambda_s$, the ring-term contribution is important and the condition $N_s \lambda_{hs} + N_\phi \lambda_{h\phi} < 0$ in \eqref{eq:constraint:neg:Vth} can be relaxed. 
%%%%%%%%%%%%%%%%
\begin{table}[bh!]
\centering
  \renewcommand{\arraystretch}{1.5}
      \addtolength{\tabcolsep}{0.2pt} 
\scalebox{0.95}{
\begin{tabular}{|c|c|c|c|c|c|c|c|c|c|c|c|c|c|}
\hline
 & $\lambda_{hs}$ & $\lambda_{h\phi}$ & $\lambda_{s}$ & $\lambda_{\phi}$& $m_s$ &  $\alpha=\frac{m_\phi^2}{\lambda_{h\phi} v^2} $ & $N_s$ & $N_\phi$ & $S_4$ & $\left(\frac{S_3}{T}\right)_{\rm min}$ & $\frac{T_{\rm min}}{v}$ & $\frac{T_c}{v}$\\
 \hline 
1 &$-0.1$ & $0.70$ & $5\cdot 10^{-4}$ & $10^{-4}$ & $246$ & $0.001$ &  $1500$ & $100$ & $36680$ & $409$ & $0.234$ & $0.343$ \\[3pt]
\hline
2 &$-0.1$ & $0.83$ & $0.1$ & $0.1$ & $246$ & $0.001$ &  $100$ & $100$ & $879$ & $386$ & $0.095$ & $0.278$ \\
\hline
\end{tabular} 
}
\caption{Two benchmark points (dimensionful parameters are in unit of GeV and $v=246$ GeV). $\lambda_s$ was chosen to satisfy the constraint in Eq.~(\ref{eq:constraint:neg:Vth2}) for the range of $N_s$ that will be used in our numerical scan. 
}
\label{tab:benchmark}
\end{table}

%%%%%%%%%%%%%%%%%%%%%%%
%%%%%%%%%%%%%%%%%%%%%%%

To illustrate the behaviors of the thermal potentials for our model, we show their values as a function of $h$ in Fig.~\ref{fig:benchmark:Tevolution}. Here, we have shifted the thermal potentials by an $h$-independent value to have a zero potential value at $h=0$. At $T=0$, the one-loop effective potential $V_{\rm tree} + V_{\rm CW}$ (the solid blue line) shows that the vacuum $\langle h \rangle = v$ is a metastable one with the global vacuum at $\langle h \rangle = 0$. At high temperatures with $T > T_c$, the symmetry breaking vacuum has a smaller effective potential than the symmetry-conserving one. In this plot, we also show the behavior at $T = T_m < T_c$ in the solid black line when $S_3(T)/T$ has its smallest value. In the right panel for the scenario 2, a monotonic behavior for the energy difference between the two vacua can be observed when $T$ decreases. On the other hand, a non-monotonic behavior appears in the left panel for the scenario 1. As a result, the potential barrier for $T=0$ potential is effectively higher than the $T=T_m$ one, so the tunneling action value $S_4$ is much larger than $S_3(T_m)/T_m$ for this benchmark point.

%%%%%%%%%%%%%%%%%%%%%%%
\begin{figure}[tph]
\begin{center}
\includegraphics[width=0.48\textwidth]{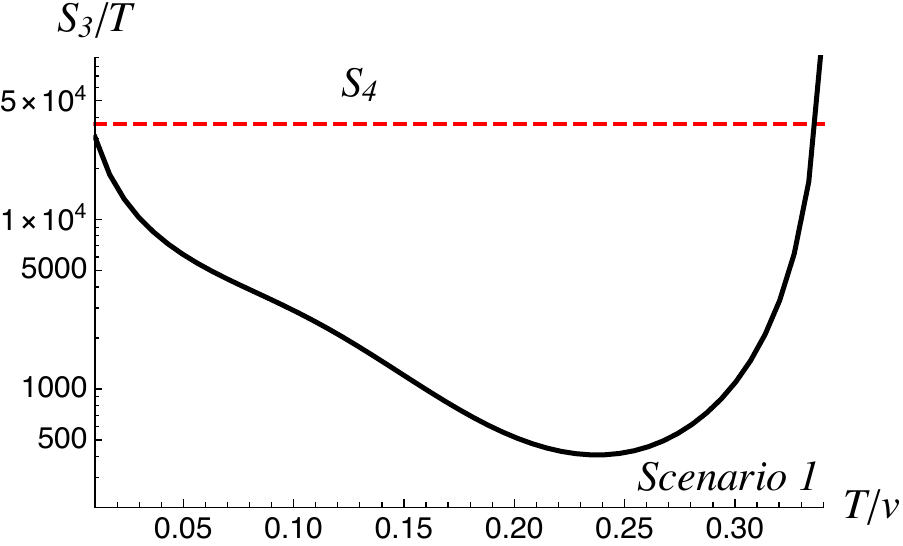} \quad
\includegraphics[width=0.48\textwidth]{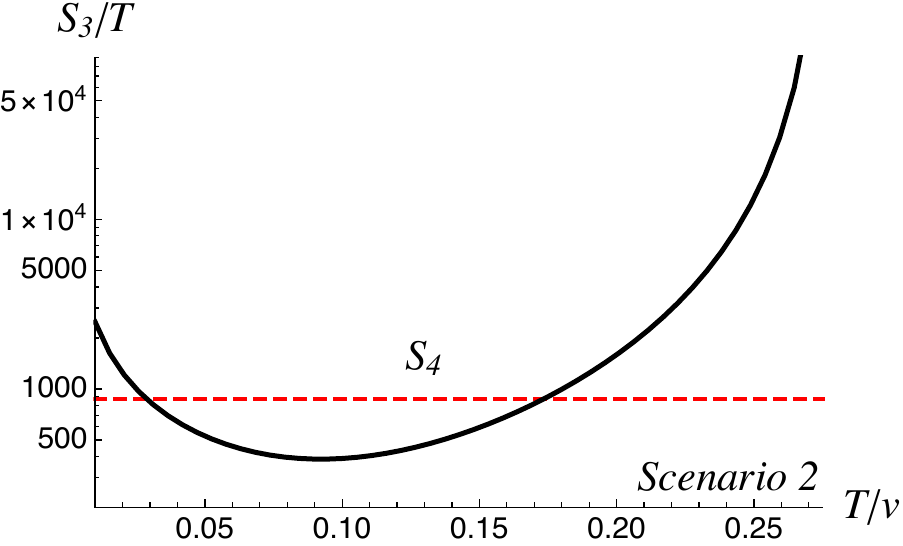}
\caption{\small $S_3(T)/T$ (solid black) as a function of $T/v$ for the benchmark scenario 1 (left) and 2 (right) in Table~\ref{tab:benchmark}. The dashed red line corresponds to the $S_4$ value for each scenario. The upper end in the $T/v$ axis corresponds to the critical temperature $T_c/v$.  
}
\label{fig:benchmark:o3bounce}
\end{center}
\end{figure}
%%%%%%%%%%%%%%%%%%%%%%%

In Fig.~\ref{fig:benchmark:o3bounce}, we show the values of $S_3(T)/T$ as a function of $T$ for both benchmark points in Table~\ref{tab:benchmark}. The increasing behavior at the high-$T$ end is due to that  $S_3(T) \propto (T_c - T)^{-a}$ when $T\rightarrow T_c$. The positive numeric index $a$ is not important for our discussion here. The increasing behavior at the low-$T$ end is simply due to that $S_3(T)$ reaches a constant such that $S_3(T)/T \propto 1/T$ when $T \rightarrow 0$. For both benchmark points, the minimum value of $S_3(T)/T$ is smaller than $S_4$. Therefore, when we consider the stability of the metastable vacuum, we have to take into account the potentially faster thermal tunneling rates.

%%%%%%%%%%%%%%%%%%%%%%%
\begin{figure}[th!]
\begin{center}
\includegraphics[width=0.48\textwidth]{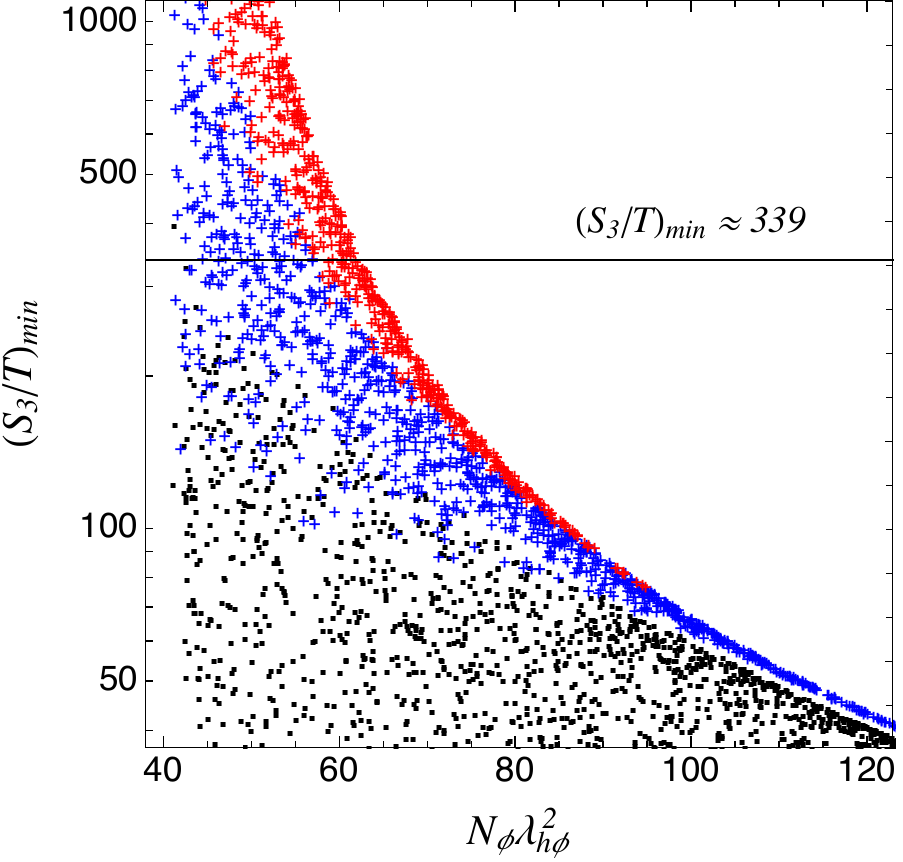} \quad
\includegraphics[width=0.48\textwidth]{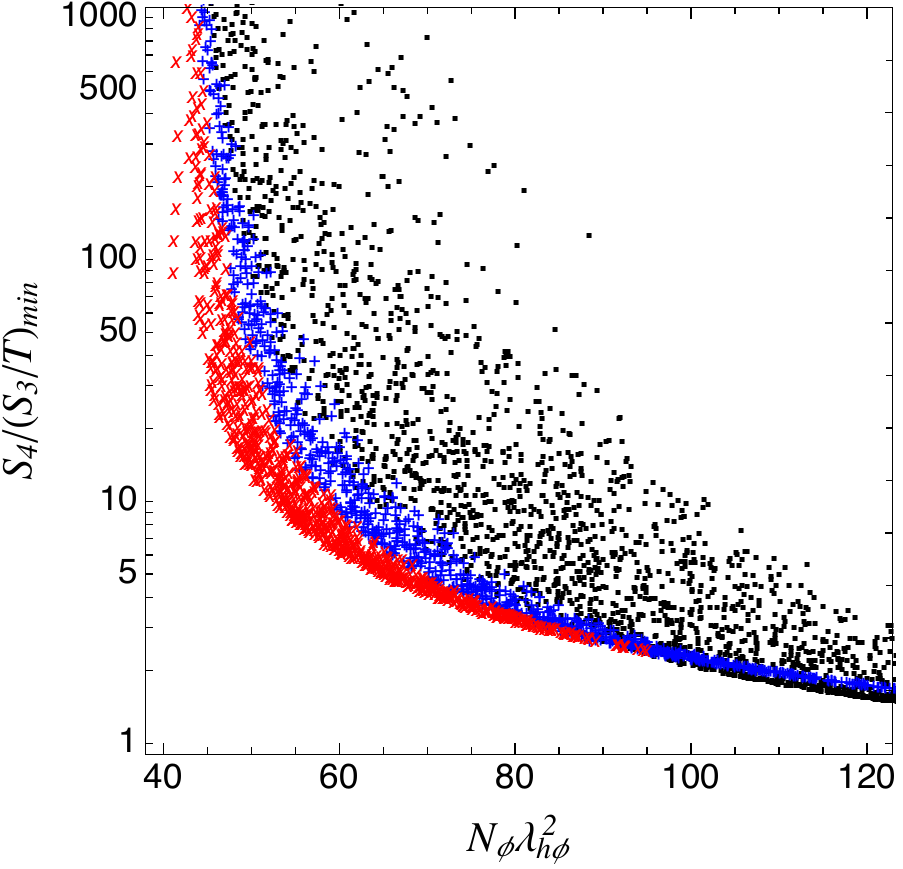} \\
\includegraphics[width=0.48\textwidth]{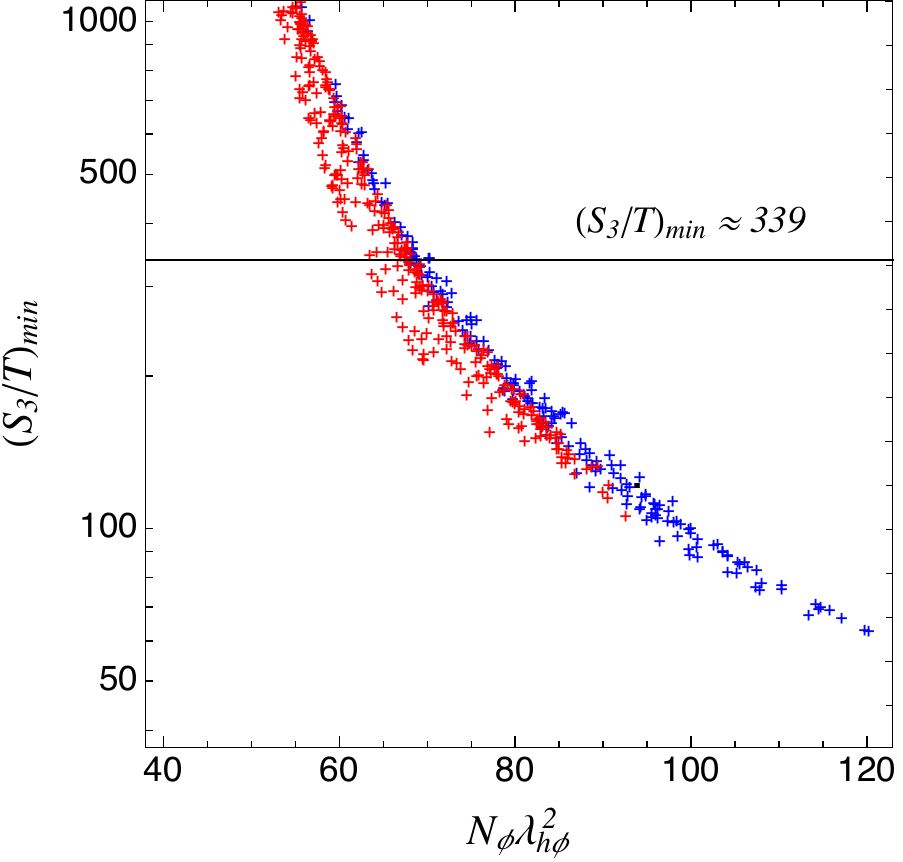} \quad 
\includegraphics[width=0.48\textwidth]{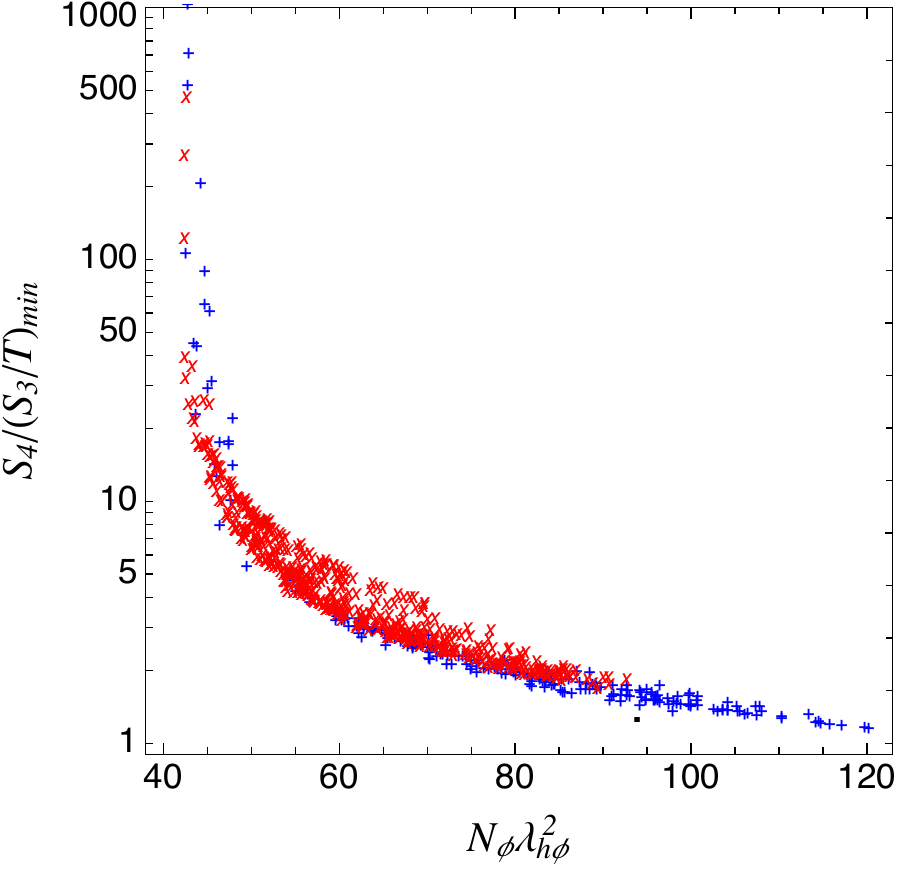}
\caption{\small $(S_3/T)_{\rm min}$ (left) and the ratio of $S_4/(S_3/T)_{\rm min}$ (right) as a function of $N_\phi \lambda^2_{h\phi}$. Upper (bottom) plots are obtained by scanning parameters around the benchmark scenario 1 (scenario 2) in Table~\ref{tab:benchmark}. Data points are subdivided into three sub-regions: $N_\phi=[30,\, 80]$ (red), $N_\phi=[80,\, 110]$ (blue) and $N_\phi=[110,\, 200]$ (black). The horizontal line is to require an uncompleted thermal phase transition from the metastable vacuum $\langle h \rangle = v$ to the global one or $S_3/T > (S_3/T)_{\rm min}\approx 339$, which infers that $N_\phi \lambda^2_{h\phi} \lesssim 62\, (69)$ from upper (bottom) plots. 
}
\label{fig:o3bounce:1D}
\end{center}
\end{figure}
%%%%%%%%%%%%%%%%%%%%%%%

To have a broader view of the values of $S_3(T)/T$ and $S_4$ for our model, we make some scans of model parameters in the vicinity of the two benchmark scenarios in Table~\ref{tab:benchmark} that have different importance for the ring terms. 
Around the first benchmark scenario,  we vary $\lambda_{h\phi}$, $N_\phi$, and $N_s$ in the following range, 
\begin{equation}\label{eq:scan:sc1}
  \lambda_{h\phi} = [0.40,\, 1.10]~, \quad
  N_\phi = [30,\, 200]~,\quad
  N_s = [1000,\, 1500]~,
\end{equation}
while fixing $\lambda_{hs}$, $\lambda_{s}$, $\lambda_\phi$, $\alpha$  (or $m_\phi$), and $m_s$ as the values in Table~\ref{tab:benchmark}.
Similarly around the second benchmark scenario,  we vary $\lambda_{h\phi}$, $N_\phi$, and $N_s$ in the following range, 
\begin{equation}\label{eq:scan:sc2}
  \lambda_{h\phi} = [0.40,\, 1.10]~, \quad
  N_\phi = [30,\, 200]~,\quad
  N_s = [30,\, 200]~,
\end{equation}
while fixing other parameters as the values in Table~\ref{tab:benchmark}. We collect only the parameter points as valid ones that exhibit non-vanishing values of $T_c$, $S_4$, and $(S_3/T)_{\rm min}$ simultaneously.
Since the parameter combination $N_\phi \lambda_{h\phi}^2$ is the most important one to have a global EW symmetric vacuum, we show $(S_3/T)_{\rm min} \equiv S_3(T_m)/T_m$ as a function of $N_\phi \lambda_{h\phi}^2$ in the left panels of Fig.~\ref{fig:o3bounce:1D}. Because $S_3/T$ depends also on the parameter combination $N_\phi \lambda_{h\phi}$ (see \eqref{eq:Vth:toy:highT}), the value of $(S_3/T)_{\rm min}$ is larger for a smaller value of $N_\phi$ (or equivalently a larger value of $\lambda_{h\phi}$) when $N_\phi \lambda_{h\phi}^2$ is fixed. To have $(S_3/T)_{\rm min} > 339$ or to have the current universe still sit at the $\langle h \rangle = v$ vacuum, one has an approximately upper bound on $N_\phi \lambda_{h\phi}^2$ as $N_\phi \lambda_{h\phi}^2 \lesssim 62\, (69)$ for the benchmark scenario 1 (scenario 2). Varying the bare mass of $S$ in the range $m_s = [200,\, 350]$ GeV in addition to the parameters in Eq.~(\ref{eq:scan:sc1})  ($m_s = [200,\, 350]$ GeV, $\lambda_\phi = [0.08,\, 0.12]$, and $\lambda_s = [0.08,\, 0.12]$ in addition to those in Eq.~(\ref{eq:scan:sc2})) for the benchmark scenario 1 (scenario 2), we have found that the upper limit remains almost the same within 3\% (1\%) upward fluctuation. Therefore, the uncompleted thermal phase transition requires 
\beqa
\mbox{\it  uncompleted thermal phase transition:}\quad\quad 41 < N_\phi \lambda_{h\phi}^2 \lesssim 70 ~.
\eeqa
In the right panels of Fig.~\ref{fig:o3bounce:1D}, the ratio of $S_4$ over $(S_3/T)_{\rm min}$ is shown as a function of $N_\phi \lambda_{h\phi}^2$. One can clearly see that this ratio is much larger than one for the cosmologically viable parameter region $N_\phi \lambda_{h\phi}^2 \lesssim 62\, (70)$. Thus the quantum tunneling at the current era is less efficient compared to the thermal tunneling in the early universe.

%%%%%%%%%%%%%%%%%%%%%%%%%
% AdS Bubble
%%%%%%%%%%%%%%%%%%%%%%%%%
\section{Formation and dynamics of bubbles}
\label{sec:bubble}
During the thermal evolution, the Higgs potential energy in the symmetry breaking phase becomes higher than the value at the origin when the temperature is below the critical temperature $T_c$ (see Fig.~\ref{fig:benchmark:Tevolution} for illustration). Since then, the vacuum at the origin becomes the global one, and the transition from the metastable symmetry breaking vacuum to the global one can happen through a first order phase transition. Bubbles with an unbroken EW symmetry inside and symmetry breaking outside could be thermally generated in space. At a low temperature such that the thermal energy is smaller than the vacuum energy difference $\epsilon(v)$ in Eq.~\eqref{eq:metavac:cons2}, the space-time geometry of the bubbles are dominated by the negative vacuum energy (our current universe has a tiny vacuum energy), those bubbles become AdS bubbles. On the other hand, to the outside observer,  the bubbles including the AdS bubbles can be treated as normal matter whose mass gets two contributions from both the volume and surface energy terms~\cite{Espinosa:2015qea}. Depending on the size of bubble radii, bubbles are sub-divided into three categories: subcritical, critical and supercritical bubbles.

The dynamical evolution of the bubbles with a global vacuum inside might be analyzed following a similar strategy as Ref.~\cite{Espinosa:2015qea} which exploits the equation of motion on the bubble wall from the metric continuity conditions at the boundary separating inside and outside of the bubbles. However, the qualitative behaviors of different kinds of bubbles can be understood using the thin-wall approximation (also the Newtonian limit applies to small-radius bubbles), although some of our model parameters have a quantitatively thick-wall behavior, which we rely on numerical calculations. 
Within the thin-wall approximation, the $O(3)$-symmetrical finite-temperature action can be approximated by the volume vacuum energy and the surface tension energy terms~\cite{Coleman:1977py,Coleman:1980aw},
\begin{equation}\label{eq:S3:thinwall}
  S_3 = 4\pi \int_0^\infty dr\, r^2 \Big [ \frac{1}{2} \left ( \frac{dh}{dr} \right )^2 + V(h,\, T) \Big ]
  \sim - \frac{4}{3} \pi R^3 \epsilon(T) + 4\pi R^2 S_1(T)~,
\end{equation}
where $r=|\vec{x}|$ and $S_1(T)$ is defined as
\begin{equation}
  S_1(T) = \int_0^\infty dr\, \Big [ \frac{1}{2} \left ( \frac{dh}{dr} \right )^2 + V(h,\, T) \Big ]~,
\end{equation}
after substituting the bounce solution for $h$. 
The critical bubble radius can be obtained by extremizing $S_3$ of Eq.~\eqref{eq:S3:thinwall} in $R$ and is~\footnote{This estimate parametrically agrees with the expression in the Newtonian limit from the alternative approach in~\cite{Espinosa:2015qea} that does not assume a thin-wall approximation.}
\begin{equation}
R_{\rm crit} \equiv \frac{2\,S_1(T)}{\epsilon(T)} ~.
\end{equation}
The bubbles with larger (smaller) radius than $R_{\rm crit}$ are classified as supercritical (subcritical) bubbles. For those supercritical bubbles with $R > R_{\rm crit}$, the volume contribution in Eq.~(\ref{eq:S3:thinwall}) dominates and it will expand. On the other hand, subcritical bubbles with $R < R_{\rm crit}$ will eventually shrink and collapse as the surface (or tension) contribution in Eq.~(\ref{eq:S3:thinwall}) dominates.  We discuss the consequence of both types of bubbles in sequence.

%%%%%%%%%%%%%%%%%%%%%
%%%%%%%%%%%%%%%%%%%%%
\subsection{Supercritical bubbles and uncompleted phase transition}
\label{sec:supercritical}

For the supercritical bubbles, they will keep expanding after productions until they meet other bubbles and percolate. If they exist within the Hubble patch of the current visible universe, their radii will be enormous. The enclosed large amount of vacuum energy makes them behave as very heavy matter and likely to overclose the universe. Assuming a constant wall velocity of $v_w$ and the initial production time at $T_m$, the size of a supercritical bubble today is $R_0\sim v_w\ t_0$ with the age of universe $t_0 \approx 13.7$\,Gyr. The total energy contained inside the bubble can be estimated to be $\epsilon(v)\,R_0^3 \sim v^4 \, v_w^3\,t_0^3$, which is $(v/\Lambda_{\rm CC})^4\,v_w^3 \sim 10^{56}\,v_w^3$ times the critical energy density of the current universe.  Here, the cosmological constant scale is $\Lambda_{\rm CC}\approx 2\times 10^{-3}$\,eV. Obviously, this is phenomenologically unacceptable. Therefore, we will require a small probability to have one supercritical bubble in the current universe, which sets a lower bound on $S_3(T)/T$.

Given the nucleation rate per volume $\Gamma_3$ in Eq.~(\ref{eq:toy:S3vsS4}), the fraction of space for a volume $V$ not occupied by the bubbles is roughly $f(t) \sim \exp (- \Gamma_3 \cdot V)$~\cite{Guth:1981uk}
\begin{equation}\label{eq:frac:not:occupied}
  f(t) = \exp \left ( - \frac{4\pi}{3} \int_{t_c}^{t} dt' v_w^3 (t-t')^3 \,\Gamma_3(t') \right )~.
\end{equation}
Since the nucleation rate is dominated by around the time with $T=T_m$ with $S_3/T$ taking its minimum value $(S_3/T)_{\rm min}$, one can derive a lower bound on the bounce action by requiring $f(t_0) \sim 1$ in the current universe (see Appendix~\ref{app:sec:frac:bubbles} for a more detailed derivation)~\footnote{In a typical EWPT, $S_3/T \sim 150$ leads to an order-one fraction of universe by bubbles and ending of the EW phase transition. For our case, the phase transition is not completed. The difference between 150 and 339 amounts to the time interval between roughly the electroweak phase transition time and today. A higher value of $S_3/T$ for our case is required to avoid even a single supercritical bubble in the current universe.}
\begin{equation}\label{eq:lowerbound:S3T}
 \frac{S_3(T_m)}{T_m}  \gtrsim \ln  \left(M_{\rm pl} T^2_m t_0^3 \right) + \frac{3}{2} \ln \left[ \frac{2}{3} \ln (M_{\rm pl} T^2_m t_0^3)\right]\approx 339~,
\end{equation}
where we have used $M_{\rm pl} = 2.43 \times 10^{18}$ GeV, $T_m = 10$ GeV and $t_0 \approx 13.7$ Gyr.

%%%%%%%%%%%%%%%%%%%%%
%%%%%%%%%%%%%%%%%%%%%
\subsection{Subcritical bubbles}

For subcritical bubbles, they have smaller free energy that can be seen from the bounce action $S_3(T)$ in Eq.~(\ref{eq:S3:thinwall}). Therefore, they are produced more frequently and may have some non-trivial cosmological consequence (see Refs.~\cite{Gleiser:1991rf,Gleiser:1995er} for the effects of subcritical bubbles on first-order phase transition). On the other hand, the subcritical bubbles prefer to shrink once generated. Hence, usually we don't anticipate any footprints they leave. Nevertheless, we discuss a few potential phenomenological consequences that they may induce. 

The first possibility is that the subcritical bubbles may collide with each other and generate some stochastic gravitational  waves. We want to show here that the energy contained in gravitational waves is unlikely to be enough to be observed. The first argument relies on the typical distance between two subcritical bubbles (or the average inter-bubble distance), which is larger than $R_{\rm crit}$ and hence the radius of subcritical bubble ($< R_{\rm crit}$). 
It can be estimated from the number density of bubbles, $n_{\text{bubble}} = N_{\text{bubble}}/V \equiv (1/d_\star)^3$. 
Since the number density can be separately estimated in terms of the nucleation rate $\Gamma(t)$ and the fraction $f(t)$~\cite{Guth:1981uk}, the typical distance scale $d_\star$ is given by
\begin{eqnarray}
d_\star \sim n_{\text{bubbles}}^{-1/3} \sim \left[\int_{t_\star}^{t}dt' \, \Gamma(t') \, f(t') \right]^{-1/3},
\end{eqnarray}
where $t_\star$ is the moment when a subcritical bubble was created and $t$ must be within the lifetime of the subcritical bubble. 
Treating the universe as a radiation-dominated one, our conservative estimate has $d_\star \geq \mathcal O(10-10^2)/v$ for our benchmark points which looks comparable to the size of those bubbles during their lifetime (see Appendix~\ref{app:sec:interbubble:dist} for details). In a more realistic situation, we expect that average inter-bubble distance is much bigger than the size of the bubbles.
The second argument is based on the smallness of energy contained in the subcritical bubbles (for supercritical bubbles, the bubble radius is large when two bubbles expand first and collide, so the energy contained in the bubble is large). The total mass contained in the subcritical bubble is small: $M \sim (4\pi/3)R^3 v^4 \sim v$, so the gravitational wave amplitude from two bubble collision is suppressed by $G_N M/R \sim v^2/M_{\rm pl}^2$. 

The second possibility is to have additional matter to stop the subcritical bubbles from shrinking. If there is a conserved quantum number like the baryon number or dark matter number (one can easily make $\Phi$ in our model a complex field to have an additional unbroken $U(1)_\Phi$). The quantum pressure from either fermions or scalar bosons can balance the surface tension pressure and make the whole system stable. This type of objects is similar to the $Q$-ball in the literature~\cite{Friedberg:1976me,Coleman:1985ki} (see also \cite{Ponton:2019hux} for the electroweak symmetric dark matter balls). For fermionic $Q$-balls, one could supplement Eq.~\eqref{eq:S3:thinwall} with the fermion kinetic energy $\sim N^{4/3}/R$ for the total fermion number of $N$. Minimizing the energy, one has the equilibrium radius of $R_{\rm eq} \sim N^{4/9}/S_1^{1/3}$ and the total mass of $M_N \sim N^{8/9}\,S_1^{1/3}$. So, for fermions with their masses mainly from the Higgs VEV, the $Q$-ball-like state with zero fermion mass inside has a smaller mass per quantum number $M_N / N \propto N^{-1/9}$ for a sufficiently larger $N$. So, the subcritical bubbles provide us another way to form $Q$-balls in the early universe, which could serve as another interesting dark matter state. The detailed properties of those states depend on the chemical potential or dark matter asymmetry and the later evolution of objects with different quantum numbers.  

The third interesting possibility is to use the subcritical bubbles to provide the out-of-equilibrium condition for the EW baryogenesis. Note that for the ordinary EW baryogenesis based on the first-order phase transition from the EW symmetric one to the broken one, the out-of-equilibrium condition is provided by the expanding supercritical bubbles. The EW sphaleron process inside the subcritical bubbles can be efficient to generate baryon number asymmetry because of the EW unbroken core. Similar to the ordinary case, additional CP-violating interactions are required and beyond the ingredients in our current model. We leave this possibility to future explorations.

%%%%%%%%%%%%%%%%%%%%%%%%
% Phenomenology
%%%%%%%%%%%%%%%%%%%%%%%%
%\section{Phenomenological implication}
\section{Collider tests of the global EWS vacuum}
\label{sec:pheno}
The particle-physics phenomenology of our scenario is similar to other Higgs-portal models with new light scalars but with a large multiplicity. To avoid the constraints from the exotic or invisible Higgs decays (assuming both $\Phi$ and $S$ are stable at the collider scale), we restrict our consideration to the case with $m^{\rm phys}_{s,\, \phi}  > m_h/2$ with $m^{\rm phys\,2}_s \equiv m^2_s+ \lambda_{hs} v^2$ and $m^{\rm phys\,2}_\phi \equiv m^2_\phi+ \lambda_{h\phi} v^2$. Otherwise, one has a too large invisible Higgs decay width. For instance, using $\Gamma(h \rightarrow \Phi \Phi) =N_\phi\, \lambda_{h\phi}^2 v^2/(8\pi m_h)$ by ignoring the phase space factor, one has the invisible Higgs decay width much larger than the total Higgs boson decay width even for $N_\phi\, \lambda_{h\phi}^2$ order of unit. Given the large multiplicity and sizable portal couplings, one could infer the existence of new scalars either via indirect constraints from the SM Higgs properties or direct production from off-shell Higgs decays.  In this section, we will consider relevant existing constraints and discuss the prospects at future experiments.

\subsection{Higgs self-couplings}
The potential existence of a global vacuum at $\langle h\rangle = 0$ brings sizable changes for the local Higgs potential shape around $\langle h\rangle = v$, which can be easily seen from Fig.~\ref{fig:outline}. So, measuring the Higgs boson self-interacting couplings will be a promising way to probe the scenarios presented in this paper. 
Defining the quantum field $\mathtt{h}\equiv h - v$ around the EW symmetry breaking vacuum, the generic Higgs potential up to the quartic coupling can be parametrized as 
\begin{equation}
 V ( \mathtt{h}) = \frac{1}{2}m^2_h\, \mathtt{h}^2 + \frac{1}{3!}\lambda_3\, v\, \mathtt{h}^3 + \frac{1}{4!} \lambda_4\,   \mathtt{h}^4~,
\end{equation}
where the SM values of Higgs self-couplings are given by $\lambda_3^{\rm SM} = \lambda_4^{\rm SM}  = 6\lambda_h = 3m^2_h/v^2$. 

Based on the total potential $V_{\rm eff} \equiv V_{\rm tree} + V_{\rm CW}$ and the CW potential in \eqref{eq:Veff:toy}, one can derive the cubic and quartic couplings in our model as
\beqa
\lambda_3 &\approx& 6 \, \lambda_h \,+\, \frac{N_\phi\,\lambda_{h\phi}^2}{4\pi^2} \,+\, \frac{N_s\,\lambda_{hs}^2}{4\pi^2}\,\frac{\lambda_{hs}\,v^2}{m_s^2} ~, \\
\lambda_4 &\approx&6 \, \lambda_h \,+\, \frac{N_\phi\,\lambda_{h\phi}^2}{\pi^2} \,+\, \frac{3\,N_s\,\lambda_{hs}^2}{2\pi^2}\,\frac{\lambda_{hs}\,v^2}{m_s^2} ~,
\eeqa
keeping the leading term in $\lambda_{hs} v^2/m_s^2$ for $\lambda_{hs} v^2 \ll m_s^2$. Ignoring the small last terms in the above two formulas, it is evident that there are positive shifts for both couplings from new physics. This can be easily seen from the cubic term $\frac{1}{3!}\lambda_3\, v\,\mathtt{h}^3  = \frac{1}{3!}\lambda_3\,v \,(h -v)^3$. To have a lower potential value at $h=0$, one prefers to have a larger positive value for $\lambda_3$. Applying the constraints, $41 < N_\phi \lambda_{h \phi}^2 \lesssim 80\,(70)$, from requiring the existence of global EWS vacuum and a long-lived current universe (uncompleted thermal phase transition), one has 
\beqa
\label{eq:lambda34}
2.3 < \lambda_3/\lambda_3^{\rm SM} \lesssim 3.6 \,(3.3) ~, \qquad   \qquad 6.3 < \lambda_4/\lambda_4^{\rm SM} \lesssim 11.4 \, (10.1) ~.
\eeqa
For the Higgs boson as the only order parameter and a global vacuum at origin, a large modification for the Higgs self-interacting cubic and quartic couplings is a robust prediction. The existing data at the LHC~\cite{Sirunyan:2018ayu,Aad:2019uzh,DiMicco:2019ngk,ATL-PHYS-PUB-2019-009,CMS-PAS-HIG-19-005} have not reached the sensitivity to probe the region in Eq.~\eqref{eq:lambda34}. For instance, using 36.1 fb$^{-1}$ data, the ATLAS collaboration has imposed an upper bound [95\% confidence level (CL)] on the cubic coupling $\lambda_3/\lambda_3^{\rm SM} < 12.0$~\cite{Aad:2019uzh}.

\begin{figure}[t!]
\begin{center}
\includegraphics[width=0.46\textwidth]{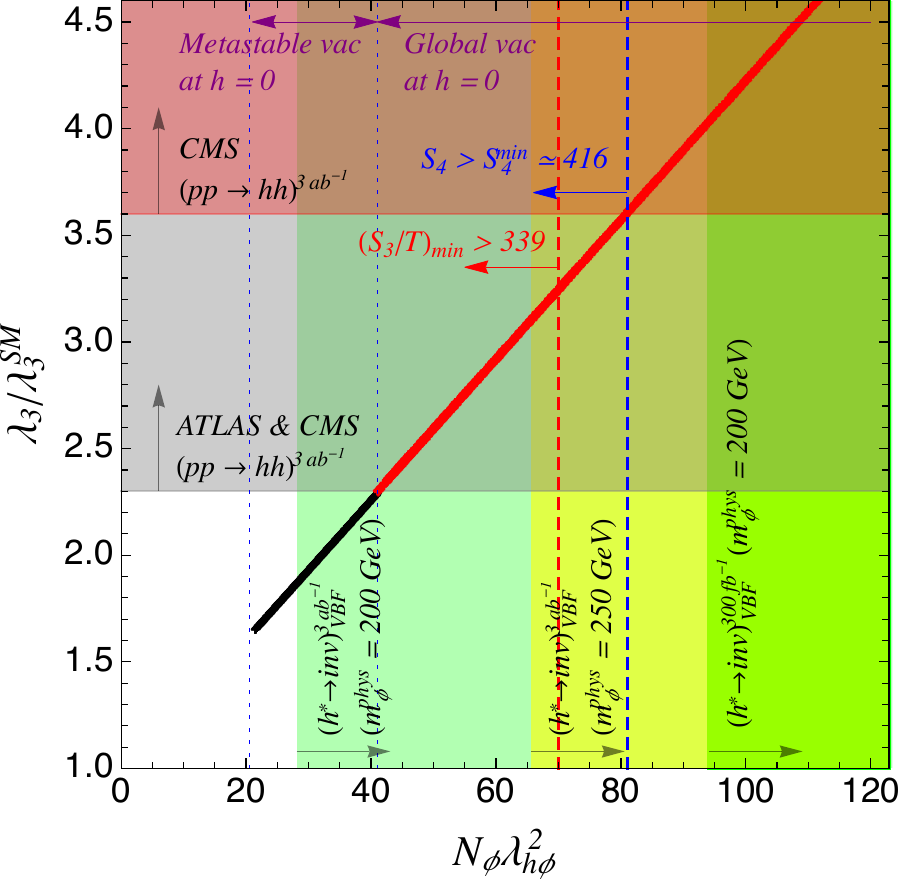} \quad
\includegraphics[width=0.46\textwidth]{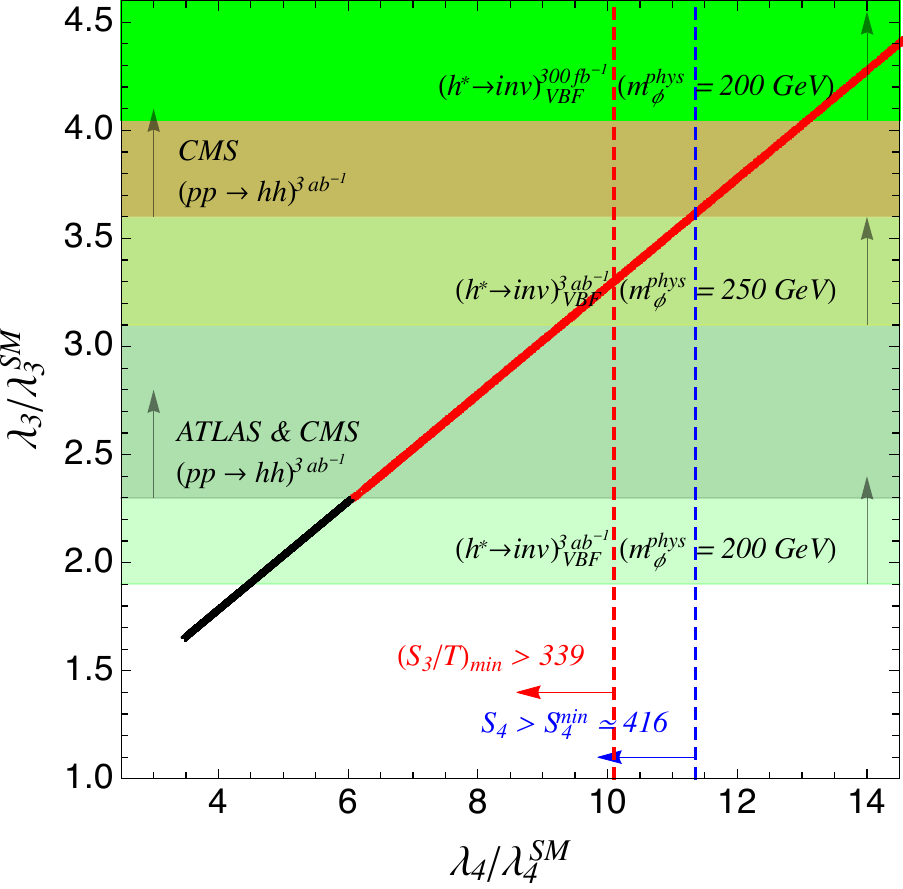} \\[20pt]
\includegraphics[width=0.46\textwidth]{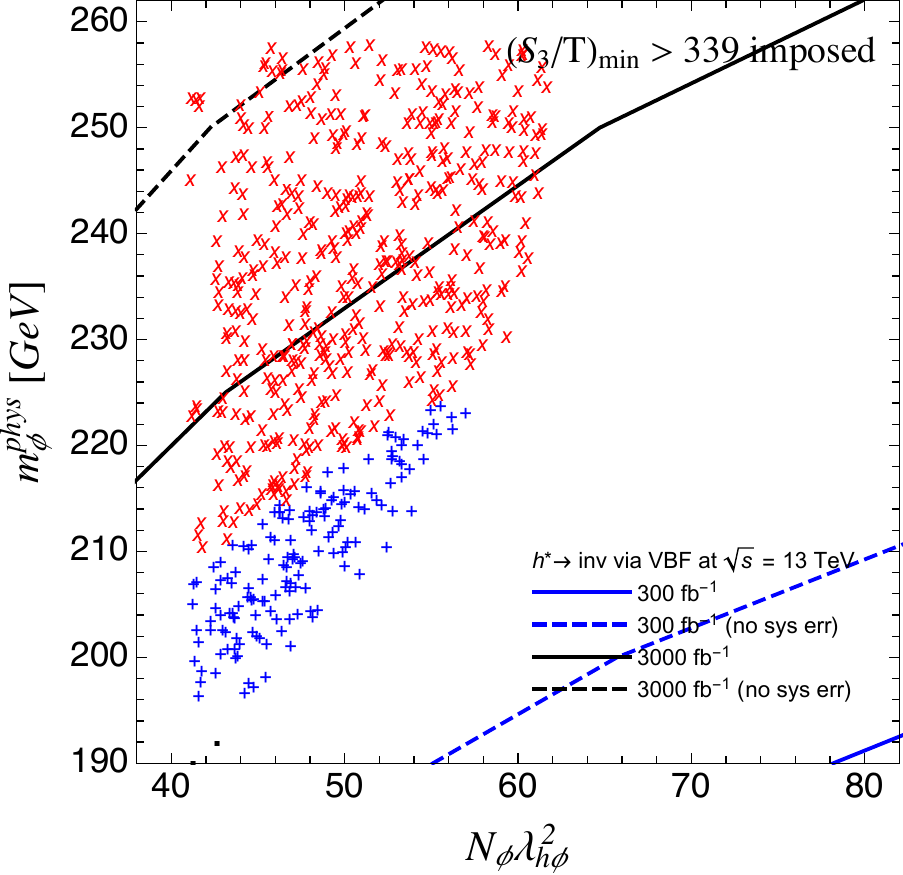} \quad
\includegraphics[width=0.46\textwidth]{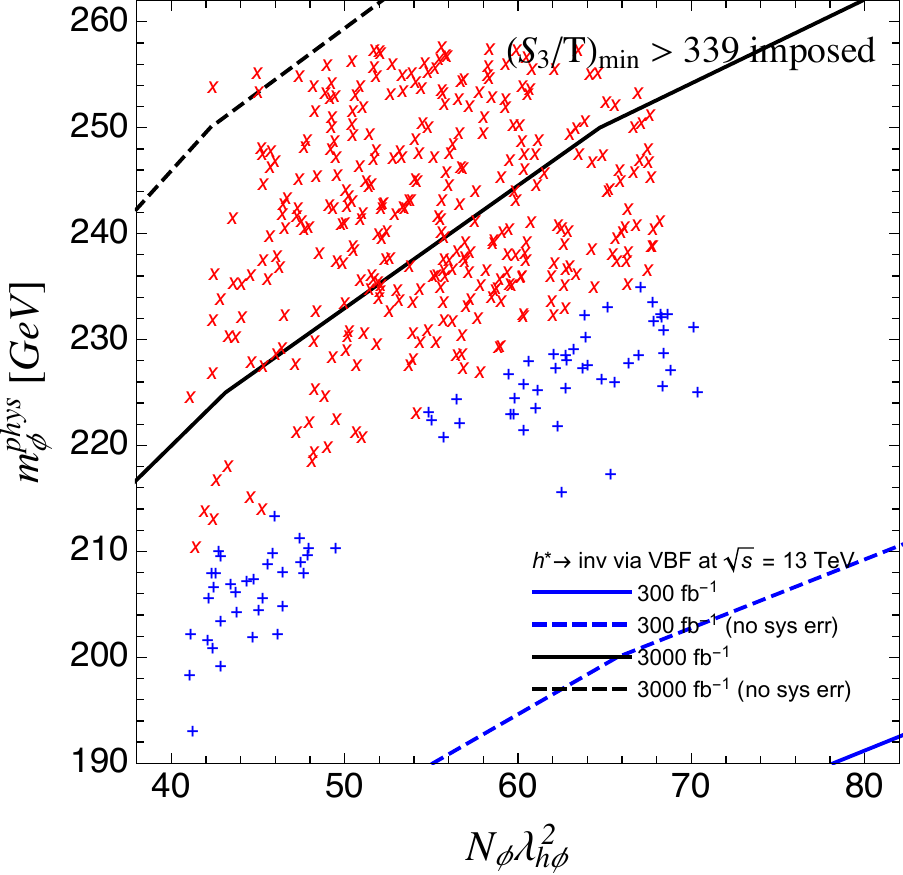}
\caption{\small Upper: Higgs cubic self-coupling, normalized to the SM value, $\lambda_3/\lambda_3^{\rm SM}$ as a function of $N_\phi \lambda^2_{h\phi}$ (left).  Predicted Higgs self-couplings in ($\lambda_3/\lambda_3^{\rm SM},\, \lambda_4/\lambda_4^{\rm SM}$) plane (right). Limits from $h^*\rightarrow {\rm inv}$ via VBF are obtained by projecting the ATLAS analysis~\cite{ATLAS-CONF-2020-008} with 139 fb$^{-1}$ data at $\sqrt{s} = 13$ TeV to higher luminosities of 300 fb$^{-1}$ and 3000 fb$^{-1}$, assuming that statistical and systematic uncertainties scale down as the square root of the luminosity. Bottom: the scattering plots for the $\Phi$ particle physical mass in terms of $N_\phi \lambda^2_{h\phi}$ for the parameter space satisfying $(S_3/T)_{\rm min} > 339$ around the benchmark scenario 1 (left) and scenario 2 (right). The same color coding as Fig.~\ref{fig:o3bounce:1D}. The projected limits from $h^*\rightarrow {\rm inv}$ via VBF are overlaid. Dashed lines correspond to the limits with only statistical uncertainties. All limits are at the 95\% CL.
}
\label{fig:bench1:c3c4}
\end{center}
\end{figure}

The HL-LHC, on the other hand, has promising potential to test the global EWS vacuum scenarios. The combined sensitivities at the 95\% CL of various decay channels of $pp\rightarrow hh$ process at the HL-LHC are $-0.4 \lesssim \lambda_3/\lambda_3^{\rm SM} \lesssim 7.3$ and $-0.18 \lesssim \lambda_3/\lambda_3^{\rm SM} \lesssim 3.6$ for ATLAS and CMS analyses, respectively~\cite{Cepeda:2019klc,DiMicco:2019ngk}. 
A simple statistical combination of ATLAS and CMS analyses, assuming no correlation among all channels, is estimated to give $0.1 \lesssim \lambda_3/\lambda_3^{\rm SM} \lesssim 2.3$ at the 95\% CL~\cite{Cepeda:2019klc,DiMicco:2019ngk}. Therefore, all or a large fraction of the global EWS vacuum scenarios will be tested at the HL-LHC (see Fig.~\ref{fig:bench1:c3c4}). The searches related to the single-Higgs production can also probe the Higgs self-couplings indirectly, although the expected constraint at the HL-LHC is weaker than that from the double-Higgs production. At lepton colliders with a low center-of-mass energy, searches associated with the single-Higgs production are the main processes to probe the Higgs self-couplings. For instance, $e^+e^-\rightarrow Zh$ where the Higgs cubic self-coupling enter through the loop correction to the $hZZ$ vertex. Recent studies at low-energy $e^+e^-$ colliders predict $\mathcal{O}(1)$ determination of the ratio $\lambda_3/\lambda_3^{\rm SM}$ at the 95\% CL. The precision becomes (or better than) roughly 50\% at the 68\% CL for the global fit (and roughly twice better for the single-parameter fit), whereas the precision at the 68\% CL at the HL-LHC has a similar precision of around 50\%~\cite{DiVita:2017vrr,DiMicco:2019ngk}. So, precise measurement of the Higgs couplings at lepton collider can also cover most of parameter space to have a global EWS vacuum. We summarize various projected sensitivities in the upper panels of Fig.~\ref{fig:bench1:c3c4} for hadron colliders.
%

%%%%%%%%%%%%%%%%%%%%%%%
%%%%%%%%%%%%%%%%%%%%%%%

Using precise measurement of the Higgs quartic self-coupling to test our scenario is generally challenging due to the tiny signal rate. It is a tough task even at the 100 TeV $pp$ collider. Recent studies on $pp \rightarrow hhh$ process at the 100 TeV $pp$ collider indicate $\mathcal{O}(10)$ determination on $\lambda_4/\lambda_4^{\rm SM}$ (for positive values), assuming 30 ab$^{-1}$ of data and the SM-like Higgs cubic self-coupling~\cite{Papaefstathiou:2015paa,Chen:2015gva,Fuks:2017zkg,Kilian:2017nio,Papaefstathiou:2019ofh} (similar sensitivities on $\lambda_4/\lambda_4^{\rm SM}$ are expected from double-Higgs production at future $e^+e^-$ or 100 TeV $pp$ colliders~\cite{Maltoni:2018ttu,Liu:2018peg,Borowka:2018pxx,Bizon:2018syu}). Nevertheless, precise measurement on the Higgs quartic coupling will be important for the model discrimination in the situation where we observe any hint of new physics based on the Higgs cubic self-coupling measurement~\cite{Jain:2017sqm}.

%%%%%%%%%%%%%%%%%%%%%%%
%%%%%%%%%%%%%%%%%%%%%%%
\subsection{Off-shell Higgs invisible decays: $E_T^{\rm miss}$+ jets}

Although there is no on-shell Higgs boson decay into the new scalars $S$ and $\Phi$ for $m^{\rm phys}_{s,\, \phi} > m_h/2$, one could search for them based on their productions at colliders from an intermediate off-shell Higgs~\cite{Craig:2014lda,Ruhdorfer:2019utl}. The production cross sections will be enhanced by the factor of $N_\phi \lambda_{h \phi}^2$ or $N_s \lambda_{h s}^2$ and could be sizable. Because of the (approximately) $O(N)$ symmetries for both scalars, $S$ and $\Phi$ are effectively stable particles at collider and appear as the missing transverse momentum ($E_T^{\rm miss}$). 

There are several production channels to probe the off-shell Higgs production of stable scalars. The process $pp \rightarrow h^*j \rightarrow SS j,\, \Phi\Phi j$ via the gluon fusion has a large cross section, although it is subject to large SM backgrounds. On the other hand, the vector boson fusion channel, $pp \rightarrow h^*jj \rightarrow SS jj,\, \Phi\Phi jj$, has the advantage of using the two forward jets to reduce the SM backgrounds. For both cases, the collider signatures are $E_T^{\rm miss} +$ jets. The production cross section of $\Phi$ (similarly for $S$) has a simple scaling for a fixed mass $m^{\rm phys}_\phi$,
\begin{equation}\label{eq:ppPhiPhi:scaling}
 \sigma(pp \rightarrow \Phi\Phi + \mbox{jets}) =  N_\phi \,\lambda_{h\phi}^2\times \sigma(pp \rightarrow h^*+ \mbox{jets} \rightarrow \Phi\Phi +\mbox{jets}: \lambda_{h\phi}\sqrt{N_\phi} =1)~.
\end{equation}
Because $N_\phi\,\lambda_{h\phi}^2 \gg N_s \lambda^2_{hs}$ for our benchmark scenarios (see Table~\ref{tab:benchmark}) and $m^{\rm phys}_\phi < m^{\rm phys}_s$, the production cross section for $\Phi$ is much larger than that for $S$. Noting that a lighter $\Phi$ particle with sizable coupling to the SM Higgs is crucial to have a global EWS vacuum at the current universe.

Using the most recent collider search for off-shell Higgs invisible decays via the VBF channel with 139 fb$^{-1}$ at ATLAS~\cite{ATLAS-CONF-2020-008},  we have simulated the signal events, imposed the corresponding kinematic cuts and found a rather weak constraint on our model parameter (see Appendix~\ref{app:sec:atlas:recasting} for more detail). For $m^{\rm phys}_\phi = 200 \, (250)$~GeV,~\footnote{The exclusion limit for different mass values is shown in Fig.~\ref{app:fig:Etjets:exclusion1} in Appendix~\ref{app:sec:atlas:recasting}.} the upper limit on $N_\phi \,\lambda_{h\phi}^2$ at the 95\% CL is $N_\phi \,\lambda_{h\phi}^2 < 137 \, (361)$. We have also made simple projections for the LHC with 300 fb$^{-1}$ and the high-luminosity LHC (HL-LHC) at $\sqrt{s} = 13$ TeV with 3 ab$^{-1}$ and found that the projected limits are $N_\phi \,\lambda_{h\phi}^2 < 94 \, (225)$ and $N_\phi \,\lambda_{h\phi}^2 < 28 \, (66)$, respectively (see Fig.~\ref{fig:bench1:c3c4} for the parameter space coverage). For the curves with systematic errors, we have assumed that the systematic errors scale down as the square root of luminosity. This could serve as a conservative estimation for the projected sensitivity.  From the upper panel of Fig.~\ref{fig:bench1:c3c4}, one can see that searching for off-shell Higgs invisible decays can probe the whole global EWS vacuum scenario at the HL-LHC if the $\Phi$ physical mass is below around 200 GeV. So, this search is complementary to measuring the Higgs cubic self-coupling to probe the Higgs potential vacuum structure. 

In the lower panels of Fig.~\ref{fig:bench1:c3c4}, we show the scattering plots of $m_\phi^{\rm phys}$ and $N_\phi \lambda^2_{h \phi}$ for the two benchmark scenarios (see Eqs.~\eqref{eq:scan:sc1} and~\eqref{eq:scan:sc2}), after imposing the condition of $(S_3/T)_{\rm min} > 339$. One can see that $m_\phi^{\rm phys}$ is above around 190 GeV for both scenarios. Noting that there is no immediate upper bound on $m_\phi^{\rm phys}$. One could reduce $N_\phi$ and hence increase $\lambda_{h \phi}$ and $m_\phi^{\rm phys}$ while keeping $N_\phi \lambda^2_{h \phi} > 41$ to have a global vacuum at $\langle h \rangle = 0$. From the projected HL-LHC sensitivities, the parameter space with $m^{\rm phys}_\phi$ below around 250 GeV can be probed with 3\,ab$^{-1}$ luminosity. Also noting that all our projected sensitivities are based on the 13 TeV LHC. Slightly better sensitivities are anticipated for the 14 TeV LHC (see Refs.~\cite{Craig:2014lda,Ruhdorfer:2019utl} for collider studies) due to roughly 20\% larger signal cross sections for the mass range of interest.

The off-shell Higgs invisible decay can also be searched for at a future lepton collider like ILC~\cite{Baer:2013cma}, CEPC~\cite{CEPCStudyGroup:2018ghi} or FCC-ee~\cite{Abada:2019zxq} if the center-of-mass energy is large enough or $\sqrt{s} > 2 m^{\rm phys}_{\phi}$ for the channel $e^+ e^- \rightarrow e^+ e^- + h^*(\rightarrow \Phi \Phi)$ and $\sqrt{s} > 2 m^{\rm phys}_{\phi} + M_Z$ for $e^+ e^- \rightarrow Z + h^*(\rightarrow \Phi \Phi)$. Depending on the collider center-of-mass energy, the projected sensitivities can exceed the ones for HL-LHC (see Ref.~\cite{Ruhdorfer:2019utl} for comparison).

%%%%%%%%%%%%%%%%%%%%%%%
%%%%%%%%%%%%%%%%%%%%%%%
\section{Discussion and conclusions}

So far, the interactions in Eq.~\eqref{eq:lag1} conserve the global $O(N_s)$ and $O(N_\phi)$ symmetries. If those global symmetries are exact, both $S$ and $\Phi$ fields are stable particles and could contribute to dark matter energy density (see \cite{Barger:2007im} for detailed studies). If the abundances of $S$ and $\Phi$ follow the thermal history and have the freeze-out ones, their abundances should be scaling like $N_s/\lambda_{h s}^2$ and $N_\phi/\lambda_{h \phi}^2$, respectively. In the large $N_s$ and $N_\phi$ limit, their thermal freeze-out abundance could be much larger than the observed dark matter energy density, and they cannot be viable dark matter candidates. On the other hand, the global symmetries could be explicitly broken by some very weak couplings such that $S$ and $\Phi$ behave as invisible particles at collider length scales. As a result, both $S$ and $\Phi$ are unstable particles and can decay into SM particles. Depending on the decay lifetime and products, they may or may not leave detectable imprints in the early universe. 

As we discussed around Eq.~\eqref{eq:landau-scale} and in Appendix~\ref{app:sec:RGE}, the effective field theory in this paper has a perturbative cutoff scale at around 100 TeV after imposing the $N_\phi \lambda^2_{h\phi}(v) > 41$ condition to have a global EWS vacuum at origin. Whether the UV theory for our model still has the global EWS vacuum depends on the detailed UV completion.  Along this direction of thought, one could also study other existing well-motivated models that address the hierarchy problem and see where one has the EWS vacuum as the global one and the EW symmetry breaking one as a metastable one. 

So far, we have found that the collider studies for either the Higgs cubic self-coupling or the VBF off-shell Higgs invisible decays can probe the majority of parameter space of having the global EWS vacuum at origin, at least at the HL-LHC with an integrated luminosity of 3\,ab$^{-1}$. The quantum tunneling at the current universe and the thermal tunneling at the early universe just provide additional constraints on the model parameter space. For instance, $S_4 \gtrsim 416$ to have a long-lived enough universe can be translated into a bound $N_\phi \lambda_{h \phi}^2 \lesssim  80$, while $(S_3/T)_{\rm min} > 339$ requires a more stringent constraint $N_\phi \lambda_{h \phi}^2 \lesssim 70$. The bubbles generated from the thermal tunneling either die off (for subcritical ones) or diluted enough for not existing within our visible Hubble patch (for supercritical ones). To have interesting small-size supercritical bubbles exist in our visible universe, the bubble growth speed needs to be dramatically suppressed and have a non-relativistic value. This could be possible if the plasma provides a strong enough friction force to reduce the bubble growth speed. We leave this possible interesting scenario with  ``electroweak symmetric bubbles in the sky" to future explorations. 

In this paper, we have only considered the scenario with the Higgs field as the only order parameter and the additional EW gauge-singlet scalars to provide one-loop potential for the Higgs field. The analysis procedure adopted here could also be applied to other scenarios with more than one order parameters to define different vacua. For those scenarios, the global EWS vacuum could be realized just by a tree-level potential with a better perturbative control. On the phenomenological side, the qualitative features of formation and dynamics of bubbles are similar to the current scenario, but the modifications to the Higgs self-couplings could be more suppressed and are more challenging to be tested at the high luminosity LHC. 

In conclusion, we have adopted a simple model to realize a global EWS vacuum at $\langle h\rangle=0$ by introducing some SM singlets with a large multiplicity to generate a substantial CW potential at loop level. We have also chosen suitable model parameters (for instance $\lambda_{hs} < 0$) to have the early universe at high temperatures sit at the EW symmetry breaking vacuum. In our study, we have focused on the case that the SM Higgs field is the only non-trivial order parameter to define different vacua with the new scalar singlets not developing a non-zero VEV. The parameter $N_\phi \lambda_{h \phi}^2$ is the most important one to achieve the global EWS vacuum if $N_\phi \lambda_{h \phi}^2 > 41$. This parameter is also bounded from above, $N_\phi \lambda_{h \phi}^2 \lesssim 70$, by requiring that the EW symmetry breaking vacuum is long-lived enough both at the current and early universe. The HL-LHC is superb to probe the scenarios discussed in this paper both from Higgs cubic self-coupling  and VBF off-shell Higgs invisible decay measurement. After recasting the existing LHC searches with 139\,fb$^{-1}$ of $pp$ collision data at $\sqrt{s} = 13$ TeV, we have found that the HL-LHC with 3\,ab$^{-1}$ can probe the parameter space with a global EWS vacuum if the $\Phi$ particle is light and below around 200 GeV.

\subsubsection*{Acknowledgments}
We would like to thank Andrew Long for useful discussion. The work of YB is supported by the U.S. Department of Energy under the contract DE-SC-0017647. MS and FY were supported by the Samsung Science and Technology Foundation under Project Number SSTF-BA1602-04. SL is supported by the National Research Foundation of Korea (NRF) grant funded by the Korea government (MEST) (No. NRF-2021R1A2C1005615).
%

%%%%%%%%%%%%%%%%%%%%%%%
%%%%%%%%%%%%%%%%%%%%%%%
%\newpage
\appendix

%%%%%%%%%%%%%%%%%%%%
%%%%%%%%%%%%%%%%%%%%
\section{Fraction of bubbles in the universe and lower bound on $S_3/T$}
\label{app:sec:frac:bubbles}
In this section, we derive the lower bound on $S_3/T$ to ensure that the vacuum of our current universe stays in the broken one during the thermal evolution of the universe.
We define $f(t)$ to be the fraction of the space not occupied by bubbles. The corresponding space will be in the electroweak symmetry breaking vacuum. Given the nucleation rate per volume $\Gamma$ (see Eq.~(\ref{eq:toy:S3vsS4})) and volume of space $V$, the fraction will be roughly given by $f(t) \sim \exp (- \Gamma \cdot V)$ where the volume $V \sim 4\pi R^3/3$ (see~\cite{Guth:1981uk} for related discussion):~\footnote{We are implicitly assuming a constant bubble wall velocity, a negligible initial bubble wall radius, and the same scale factor during the time interval of the integration.}
\begin{equation}\label{app:eq:frac:not:occupied}
  f(t) = \exp \left ( - \frac{4\pi}{3} \int_{t_c}^{t} dt' v_w^3\, (t-t')^3 \, \Gamma (t') \right )~,
\end{equation}
where $v_w$ is the bubble wall velocity.
The quantity $h(t) = 1 - f(t)$, on the other hand, will be the fraction of the space occupied by the bubbles. 
Since the transition rate $\Gamma$ has a maximum at $T \simeq T_m$, the integrand in Eq.~(\ref{app:eq:frac:not:occupied}) should be dominated by the time with the temperature $T_m$. We approximate the integrand using the saddle-point approximation:
\begin{equation}\label{app:eq:log:Gamma:saddle}
   \ln \Gamma (t') \approx \ln \Gamma(t_m) + \frac{1}{2} (t' - t_m)^2 \zeta~,
\end{equation}
where $t_m$ is the time when $\Gamma(t')$ reaches its maximum value. On the other hand, taking log of the decay rate per volume gives rise to approximately
\begin{equation}\label{app:eq:log:Gamma}
   \ln \Gamma(t') \approx \text{constant}  + \ln T^4 - \frac{S_3}{T}~,
\end{equation}
where we ignored the $\ln (S_3/T)$ term. Since $\ln \Gamma (t)$ has an extremum at $t=t_m$, its derivative using Eq.~(\ref{app:eq:log:Gamma}) should vanish
\begin{equation}
 0 = \left . \frac{d\ln \Gamma(t)}{dt} \right |_{t=t_m} \approx \left . \left [ \frac{d (-S_3/T)}{dt} - 4 H \right ]  \right |_{t=t_m} 
 = \beta (t_m ) -  \left . 4 H \right |_{t=t_m}~,
\end{equation}
where $\beta (t_m) =  \left . 4 H \right |_{t=t_m}= 2/{t_m}$ and $\beta(t)$ was defined as
\begin{equation}\label{app:eq:beta:def}
  \beta (t) \equiv \frac{d(-S_3/T)}{dt} = \frac{1}{t_m T_m^2} \frac{T^3}{2} \frac{d(S_3/T)}{dT}~.
\end{equation}
The last relation in Eq.~(\ref{app:eq:beta:def}) was obtained using $t/t_m = (T/T_m)^2$ which leads to $dt = - 2 t_m T^2_m T^{-3} dT$.
Using Eqs.~(\ref{app:eq:log:Gamma}) and (\ref{app:eq:log:Gamma:saddle}), one can express $\zeta$ as 
\begin{equation}\label{app:eq:zeta:S3T}
\begin{split}
  \zeta &= \left . \left [ \frac{d^2 (-S_3/T)}{dt^2} - 4\frac{dH}{dt} \right ] \right |_{t=t_m}
  \\[3pt]
  &=-\frac{T^3}{4 t_m^2 T^4_m} \left . \left [ 3 T^2 \frac{d (S_3/T)}{dT} + T^3 \frac{d^2 (S_3/T)}{dT^2} \right ] \right |_{T=T_m} + \frac{2}{t^2_m}~.
\end{split}
\end{equation}
We expand $S_3/T$ around $T=T_m$ where the decay rate $\Gamma$ has a maximum,
\begin{equation}\label{app:eq:S3T:series:Tm}
  \frac{S_3(T)}{T} = \frac{S_3(T_m)}{T_m} - \frac{c_1}{T_m} (T- T_m ) + \frac{c_2}{T_m^2} (T- T_m )^2~,
\end{equation}
where terms were included only up to the quadratic order. We will be interested in cases where this truncation is justified.
If we assume that $S_3/T$ has a minimum at $T= T_s$, the minimization of $S_3/T$ will relate $c_1$ and $c_2$:
\begin{equation}
  0 = \left . \frac{d(S_3/T)}{dT} \right |_{T_s} \approx - \frac{c_1}{T_m} + 2 \frac{c_2}{T^2_m} (T_s - T_m)~,
\end{equation}
which leads to
\begin{equation}\label{app:eq:c1c2:relation}
   c_1 \approx 2 c_2 \left ( \frac{T_s - T_m}{T_m} \right )~.
\end{equation}
Applying the condition on $\beta(t)$ at $t=t_m$, or $\beta(t_m) = 4\left . H \right |_{t=t_m}$ to the expansion in Eq.~(\ref{app:eq:S3T:series:Tm}) determines the coefficient $c_1$ to be
\begin{equation}
  c_1 = - 4~.
\end{equation}
Note that the origin of the non-vanishing $c_1$ is an overall $T^4$ factor in the definition of the transition rate $\Gamma(t)$.
Eq.~(\ref{app:eq:c1c2:relation}) implies that, when $|T_s - T_m | \ll T_m$, the value of $c_1$ is parametrically smaller than $c_2$, or $c_1 \ll c_2$.
In the limit of $c_1 \ll c_2$ along with the value of $c_1 = -4$, $\zeta$ in Eq.~(\ref{app:eq:zeta:S3T}), using the expression in Eq.~(\ref{app:eq:S3T:series:Tm}), becomes
\begin{equation}
   \zeta \approx - \frac{c_2}{2 t^2_m} + \frac{2}{t^2_m} \approx - \frac{c_2}{2 t^2_m} ~.
\end{equation}
Given the estimate of $\zeta$, we can evaluate approximately the integrand in the fraction $f(t)$ given in Eq.~(\ref{app:eq:frac:not:occupied}). Assuming $t_c \ll t_m \ll t$, we have
\begin{equation}
\begin{split}
\int_{t_c}^{t} dt' v_w^3 (t-t')^3 \Gamma (t') &\approx v_w^3\, \Gamma (t_m)\, t^3 \int_{t_c}^t dt' \exp \left [ -\frac{1}{2} \frac{c_2}{2 t^2_m} (t'-t_m)^2 \right ]
\\[3pt]
&\approx v_w^3\, \Gamma (t_m)\, t^3 \int_{-\infty}^{\infty} dt' \exp \left [ - \frac{1}{2} \frac{c_2}{2 t^2_m} (t'-t_m)^2 \right ]
\\[3pt]
&= v_w^3\, \Gamma (t_m)\, t^3 \sqrt{\frac{4\pi}{c_2}}\, t_m~.
\end{split}
\end{equation}
Therefore, the fraction of the space not occupied by bubbles is approximately given by
\begin{equation}
  f(t) \approx \exp \left [ - \frac{4\pi}{3} \sqrt{\frac{4\pi}{c_2}} v_w^3 t^3 t_m\, \Gamma (t_m) \right ]~.
\end{equation}
In the radiation-dominated era, the Hubble parameter is given by
\begin{equation}\label{app:eq:H:rad}
  H = \frac{1}{2t} = \sqrt{\frac{g_* \pi^2}{90}} \frac{T^2}{M_{\rm pl}} \equiv a \frac{T^2}{M_{\rm pl}} ~,
\end{equation}
where $a = \sqrt{g_* \pi^2/90}$ is taken to be an order-one number.  Using the expression in Eq.~(\ref{app:eq:H:rad}), we can convert $t_m$ into $T_m$ and the current fraction of space not occupied by bubbles reads~\footnote{If we assume the radiation-domination at any $T$ including the current temperature $T_0$, Eq.~(\ref{app:eq:H:rad}) can be rewritten as a function of temperature $T$: 
$ f(T) = \exp \biggl[ - \frac{1}{24}\frac{\sqrt{2}}{\sqrt{c_2}} \frac{v_w^3}{a^4} \frac{M^4_{\rm pl} T^2_m}{T^6} \bigl(\frac{S_3(T_m)}{T_m} \bigr)^{3/2} e^{-\frac{S_3(T_m)}{T_m}}\biggr] $. The corresponding lower bound on $S_3(T_m)/T_m$ from $f(T_0)\sim 1$ is $\frac{S_3(T_m)}{T_m}  \gtrsim \ln  \left(M_{\rm pl}^4 T^2_m /T_0^6 \right) + \frac{3}{2} \ln \left[ \frac{2}{3} \ln (M_{\rm pl}^4 T^2_m/ T_0^6]\right]\approx 358~$, where $M_{\rm pl} = 2.43 \times 10^{18}$ GeV, $T_m = 10$ GeV and $T_0 = 10^{-4}$ eV are used. This estimation is slightly larger than (\ref{app:eq:exp:lesssim:two}).}
\begin{equation}\label{app:eq:fraction:T}
  f(t_0) = \exp \left [ - \frac{1}{24}\frac{\sqrt{2}}{\sqrt{c_2}} \frac{v_w^3}{a^4} M_{\rm pl} T^2_m t_0^3 \left (\frac{S_3(T_m)}{T_m} \right )^{3/2} e^{-\frac{S_3(T_m)}{T_m}}\right ]~,
\end{equation}
where $t_0 $  is the current age of the universe.

For the evolving universe to stay in the EW symmetry breaking vacuum, the fraction of the space not occupied by bubbles at the current era must be nearly one, $f(t_0) \sim 1$. This requires that the exponent in Eq.~(\ref{app:eq:fraction:T}) should be suppressed:
\begin{equation}\label{app:eq:exp:lesssim:one}
 \text{constant}+ \ln  \left(M_{\rm pl} T^2_m t_0^3 \right) + \frac{3}{2} \ln \left (  \frac{S_3(T_m)}{T_m}  \right ) - \frac{S_3(T_m)}{T_m} \lesssim 0~ ,
\end{equation}
where the constant term and $\ln \left(S_3(T_m)/T_m\right)$ are sub-leading. Consequently, the above inequality in Eq.~(\ref{app:eq:exp:lesssim:one}) sets the lower bound on $S_3(T_m)/T_m$ as
\begin{equation}
\label{app:eq:exp:lesssim:two}
    \frac{S_3(T_m)}{T_m}  \gtrsim \ln  \left(M_{\rm pl} T^2_m t_0^3 \right) + \frac{3}{2} \ln \left[ \frac{2}{3} \ln (M_{\rm pl} T^2_m t_0^3)\right]\approx 339~,
\end{equation}
where we have used $M_{\rm pl} = 2.43 \times 10^{18}$ GeV, $T_m = 10$ GeV and $t_0 \approx 13.7$ Gyr.

%%%%%%%%%%%%%%%%%%%%
%%%%%%%%%%%%%%%%%%%%
\section{Inter-bubble distance of subcritical bubbles}
\label{app:sec:interbubble:dist}
In this section, we estimate the typical distance between subcritical bubbles  or an average inter-bubble distance. It can be estimated from the number density of bubbles, $n_{\text{bubble}} = N_{\text{bubble}}/V \equiv (1/d_\star)^3$. 
Since the number density can be separately estimated in terms of the nucleation rate $\Gamma(t)$ and the fraction $f(t)$, the typical distance scale $d_\star$ is given by
\begin{eqnarray}
d_\star \sim n_{\text{bubbles}}^{-1/3} \sim \left[\int_{t_\star}^{t}dt' \, \Gamma(t') \, f(t') \right]^{-1/3},
\end{eqnarray}
where $t_\star$ is the moment when a subcritical bubble was created and $t$ must be within the lifetime of the subcritical bubble. 
Treating the universe as a radiation-dominated one, we can rewrite the bubble number density as
\begin{eqnarray}\label{eq:nbubble:subcrit}
n_{\text{bubbles}}\sim \int _{T_\star-\Delta T} ^{T_\star} \frac{dT}{T} \frac{\Gamma(T)}{H(T)}f(T),
\end{eqnarray}
where $f(T)\to 1$ in our consideration (as a conservative estimate) and $ \Delta  T$ is the temperature change for a lifetime $\Delta \tau$ of a bubble 
\begin{eqnarray}
\Delta T \sim \sqrt{\frac{M_{\rm pl}}{2}}\left(\sqrt{\frac{1}{t_\star}} - \sqrt{\frac{1}{t_\star + \Delta \tau}}\right).
\end{eqnarray}
The lifetime of a subcritical bubble is taken to be the sum of timescales of expansion and collapse. For our benchmark points where the nucleation temperature is roughly ten times smaller than the electroweak scale, it is estimated to be $\Delta \tau \sim \mathcal O(10-10^2)/v$. It is straightforward to check that the corresponding temperature change $\Delta T$ is tiny, compared to $T_c - T_\star$ when taking $T_\star \sim 0.1 v$ [roughly the  ballpark value where $S_3(T)/T$ has a minimum] and using $t_\star \sim M_{\rm pl}/(2T_\star^2)$. 

As the conservative estimate of the typical distance scale, we approximate the integrand in Eq.~(\ref{eq:nbubble:subcrit}) with the value at $T=T_\star$. It gives rise to the lower bound on the characteristic distance scale,
\begin{equation}
\begin{split}
d_\star &\geq \left(\Delta T\, T_\star \frac{M_{\rm pl}}{(2\pi)^{3/2}}  \sqrt{\frac{90}{\pi^2 g_{*}(T_\star)}}\right)^{-1/3} \left[\left(\frac{S_3(R,\,T_\star)}{T_\star}\right)^{3/2}\,e^{-S_3(R,\,T_\star)/T_\star}\right]^{-1/3}
\\[7pt]
&\geq \left(\Delta T\, T_\star \frac{M_{\rm pl}}{(2\pi)^{3/2}}  \sqrt{\frac{90}{\pi^2 g_{*}(T_\star)}}\right)^{-1/3}\left[\left(\frac{3}{2}\right)^{3/2} \, e^{-3/2}\right]^{-1/3}~,
\end{split}
\end{equation}
where, in the second inequality, we substituted $S_3/T = 3/2$ that maximizes the transition rate to get even more conservative lower bound on the length scale.
Assuming the radiation-dominated universe, our conservative estimate gives rise to $d_\star \geq \mathcal O(10-10^2)/v$ for our benchmark 
points which looks comparable to the aforementioned lifetime. An average inter-bubble distance in a more realistic situation is expected to be much bigger than the length scale of the bubble lifetime, $d_\star \gg \Delta \tau$.

%%%%%%%%%%%%%%%%%%%%
%%%%%%%%%%%%%%%%%%%%
\section{The Landau-pole scale}
\label{app:sec:RGE}
Since new scalar fields have large multiplicities, the running couplings can rapidly change with the varying scale and it may make our calculation not trustable. 
The running of the Higgs quartic coupling in the pure SM is dominated by the negative contribution from the top Yukawa coupling at a very high scale, and the resulting instability at a large Higgs field value may hint the existence of new global vacuum. To truly establish the vacuum at the origin as the global one, one can not have an instability due to the negative Higgs quartic coupling at a large Higgs field value. In this section, we examine the evolution of some stability conditions against the running couplings as well as the Higgs quartic coupling running in presence of new light scalars.

We define the $\beta$-function for a generic coupling $g$ as
\begin{equation}
  \beta_g = \frac{dg}{d\ln \mu} = \frac{1}{(4\pi)^2} \beta_g^{(1)} + \frac{1}{(4\pi)^4} \beta_g^{(2)} + \cdots~,
\end{equation}
where $\mu$ is the renormalization scale. 
The one-loop $\beta$-functions for the gauge couplings in the $\overline{\rm MS}$ scheme are given by
\begin{equation}\label{eq:RGEGauge}
\begin{split}
\beta_{g_1}^{(1)} = \frac{41}{10} g_1^3~,\quad
\beta_{g_2}^{(1)} = -\frac{19}{6}g_2^3~, \quad 
\beta_{g_3}^{(1)} = -7 g_3^3~.
\end{split}
\end{equation}
The one-loop $\beta$-function for the top Yukawa coupling is
\begin{equation}\label{eq:RGEYukawa}
\begin{split}
\beta_{y_t}^{(1)} =&\ y_t\left(
\frac{9}{2} y_t^2 - \frac{17}{20}g_1^2 - \frac{9}{4}g_2^2 - 8g_3^2
\right)~.
\end{split}
\end{equation}
The one-loop $\beta$-functions for the scalar quartic couplings are
\begin{equation}\label{eq:RGEScalar}
\begin{split}
\beta_{\lambda_h}^{(1)} =&\ \left[ \lambda_{h}\Big ( 12 y_t^2 - \frac{9}{5}g_1^2 -9g_2^2 + 24\lambda_h \Big ) \right.
\\[3pt]
&-	\left. 
6y_t^4 +\frac{9}{20} g_1^2g_2^2 + \frac{27}{200}g_1^4 +\frac{9}{8}g_2^4 + 2 N_s \lambda_{hs}^2 + 2 N_\phi \lambda_{h\phi}^2
\right]~,
\\[3pt]
\beta_{\lambda_{hs}}^{(1)} =&\  \lambda_{hs} \Big ( 6y_t^2 -\frac{9}{10}g_1^2 -\frac{9}{2}g_2^2 +12\lambda_h + (2 N_s +4 ) \lambda_s  + 8 \lambda_{hs}  \Big )~,
\\[3pt]
\beta_{\lambda_{h\phi}}^{(1)} =&\  \lambda_{h\phi}\Big ( 6y_t^2 -\frac{9}{10}g_1^2 -\frac{9}{2}g_2^2 +12\lambda_h + (2 N_\phi + 4) \lambda_\phi  + 8 \lambda_{h\phi}  \Big )~,
\\[5pt]
\beta_{\lambda_{s}}^{(1)} =&\ 8\lambda_{hs}^2 + (2 N_s +16 ) \lambda_s^2~,
\\[5pt] 
\beta_{\lambda_{\phi}}^{(1)} =&\ 8\lambda_{h\phi}^2 + (2 N_\phi + 16 ) \lambda_\phi^2 ~.
\end{split}
\end{equation}

For qualitative understanding and because of  large $N_s$ and $N_\phi$, we can approximate one-loop $\beta$-functions by neglecting most of SM contributions.
First of all, the one-loop $\beta$-function for the Higgs quartic coupling is approximated to be 
\begin{equation}
\begin{split}
\beta_{\lambda_h}^{(1)} \approx&\   12 y_t^2 \lambda_{h} + 24\lambda_h^2 -
6y_t^4  + 2 N_s \lambda_{hs}^2 + 2 N_\phi \lambda_{h\phi}^2~,
\end{split}
\end{equation}
which shows that it receives large positive contributions from new scalar sectors and it will not develop a negative value. This excludes the possibility for a global vacuum at a large Higgs field region and it better supports that our global vacuum at the origin is the truly global one. Next, we look into the running of scalar quartic couplings. To solve the renormalization group equations (RGE's) and look for Landau pole scales, we will ignore the relatively small $\lambda_h$. The equations can be grouped into two pairs of equations. For $\lambda_{hs}$ and $\lambda_s$, one has 
\beqa
\beta_{\lambda_{hs}}^{(1)} &=&\  (2 N_s + 4)\,  \lambda_{hs} \lambda_s + 8 \, \lambda_{hs}^2  \,, \\
\beta_{\lambda_{s}}^{(1)} &=&\ 8\,\lambda_{hs}^2 + (2 N_s + 16)  \lambda_s^2 \,.
\eeqa
Given $\lambda_s > - 4 \lambda_{hs}/(N_s+2)$ from \eqref{eq:lamhphiwindow:highT} and $\lambda_{hs} < 0$ at around the electroweak scale, one has an upper bound: $\beta_{\lambda_{hs}}^{(1)} < 0$.  This means that $\lambda_{hs}$ will become more negative at a higher scale, although no direct Landau-pole scale can be obtained. 
However, if we also take into account the running of $\lambda_s$, the situation is different. At the scale of $v$, one could choose $\lambda_s \sim - \lambda_{hs}/N_s$ and ignoring the term $(2 N_s + 16) \lambda_s^2$. To solve $\lambda_s$, we first ignore the running of $\lambda_{hs}$ and have 
\beqa
\lambda_s(\mu) \approx \frac{8 \lambda^2_{hs}}{16\pi^2}\ln{(\mu/v)} ~,
\eeqa
by ignoring its small value at $v$. 
We then have 
\beqa
\beta_{\lambda_{hs}}^{(1)} &=&\  \frac{8 (2 N_s + 4)}{16\pi^2}  \lambda_{hs}^3 \ln{(\mu/v)} + 8 \lambda_{hs}^2 ~. 
\eeqa
If the first term dominants, one can integrate out the RGE to have
\beqa
\frac{1}{\lambda^2_{hs}(\mu) } = \frac{1}{\lambda^2_{hs}(v) } - \frac{8 (2 N_s + 4)}{(16\pi^2)^2}\left( \ln{(\mu/v)}\right)^2 ~.
\eeqa
The Landau pole scale is then
\beqa
\Lambda_{\rm L} = v \,\mbox{exp}\left[ \frac{16\pi^2}{\sqrt{8(2N_s+4)}}\, \frac{1}{|\lambda_{hs}(v)|} \right] ~.
\eeqa
For $N_s=1500$ and $\lambda_{hs}(v)=-0.1$, one has $\Lambda_{\rm L} = 6.5\times 10^6$~GeV. For $N_s=100$ and $\lambda_{hs}(v)=-0.1$, one has $\Lambda_{\rm L} = 2.3\times 10^{19}$~GeV. 

%%%%%%%%%%%%%%%%%%%%%%%
\begin{figure}[htb]
\begin{center}
\includegraphics[width=0.45\textwidth]{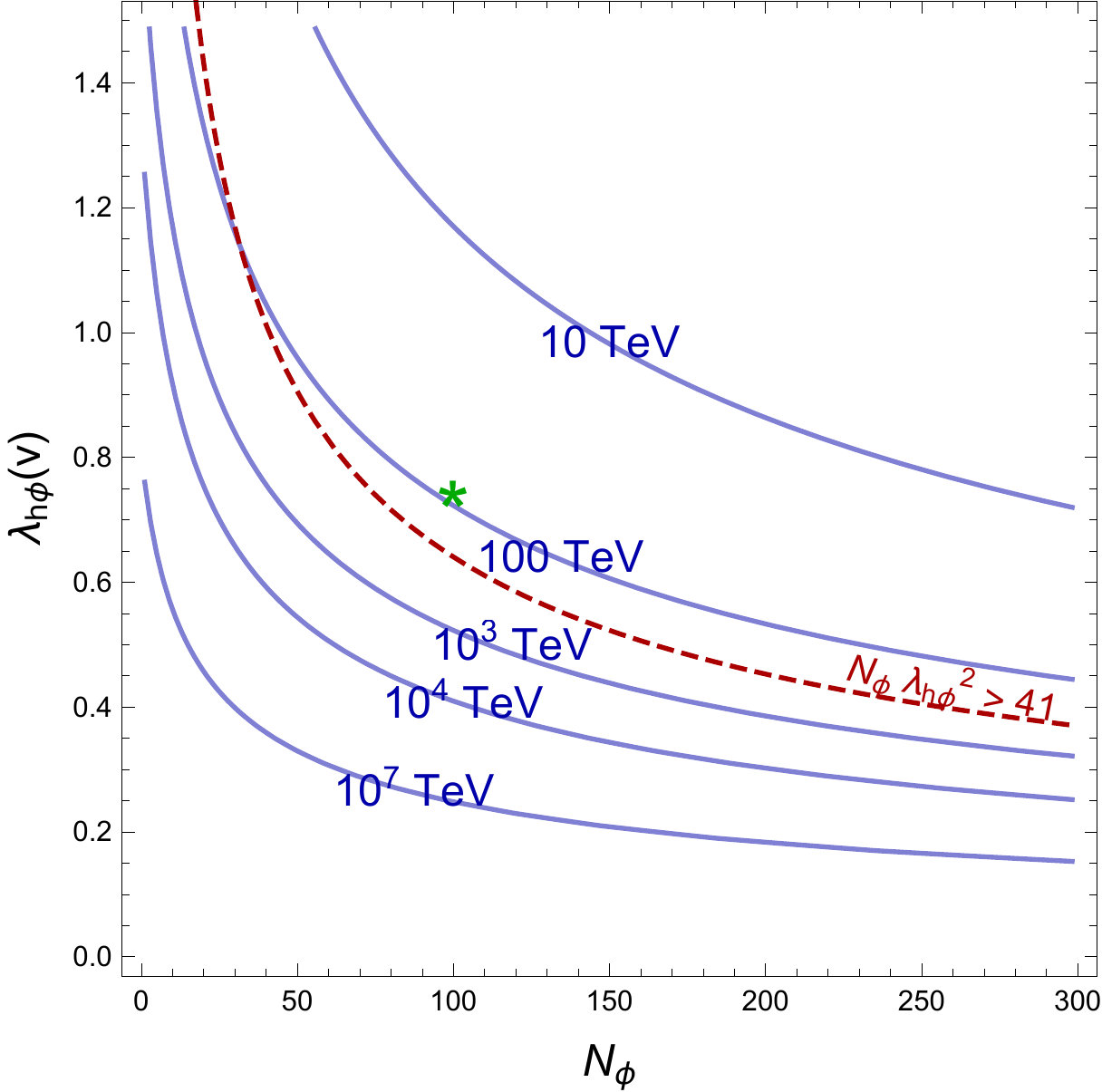}
\caption{\small The Landau-pole scales for different values of $N_\phi$ and $\lambda_{h\phi}(v)$ with $\lambda_\phi(v) = 0$. The first benchmark point in Table~\ref{tab:benchmark} is shown in the green star point. }
\label{fig:landau-pole-scale}
\end{center}
\end{figure}
%%%%%%%%%%%%%%%%%%%%%%%

A similar analysis can be applied to $\lambda_{h\phi}$ and $\lambda_\phi$. Noting the lower bound on $N_\phi\,\lambda^2_{h\phi}(v) > 41$ from \eqref{eq:constraint:pos:epsilon}. We have 
\beqa
\Lambda_{\rm L} = v \,\mbox{exp}\left[ \frac{16\pi^2}{\sqrt{8(2N_\phi+4)}}\, \frac{1}{\lambda_{h\phi}(v)} \right]  \lesssim v \,\mbox{exp}\left[ \frac{16\pi^2}{\sqrt{16\times 41}} \right] \approx 120 \,\mbox{TeV} ~, 
\eeqa
which will be the cutoff scale of our effective field theory. Above this scale, we anticipate some new particles and/or new interactions beyond the Lagrangian in the current paper. 

Numerically solving the RGEs for $\lambda_{h\phi}$ and $\lambda_\phi$, we show the Landau-pole scales for different choices of $N_\phi$ and $\lambda_{h\phi}(v)$ in Fig.~\ref{fig:landau-pole-scale} while fixing $\lambda_\phi(v) = 0$.

%%%%%%%%%%%%%%%%%%%%
%%%%%%%%%%%%%%%%%%%%
\section{Recasting ATLAS analysis of $E_T^{\rm miss} +$ jets via VBF at $\sqrt{s}=13$ TeV}
\label{app:sec:atlas:recasting}
In this section we recast the recent ATLAS analysis for invisible Higgs boson decays via the VBF productions using an integrated luminosity of 139 fb$^{-1}$ at $\sqrt{s}=13$ TeV~\cite{ATLAS-CONF-2020-008}. This invisible Higgs search leads to the same collider signature of the missing transverse momentum and two forwards jets. We have implemented our Higgs portal model in {\sc FeynRules}~\cite{Alloul:2013bka} to generate UFO model file which is then used in \textsc{\sc MadGraph}5\_aMC$@$NLO v2.3.3~\cite{Alwall:2014hca}. We have simulated our signal events using the process $pp \rightarrow h^*jj \rightarrow \Phi\Phi jj \rightarrow \nu\bar{\nu}\nu\bar{\nu}jj$ (with $\Phi \rightarrow \nu\bar{\nu}$ for modeling the missing transverse momentum) by switching on only the VBF production process with the generation level cuts, $p_T(j) > 20$ GeV and $|\eta(j)| < 5.0$. Since the multiplicity $N_\phi$ enters the cross section only through the overall scaling of $N_\phi \lambda_{h\phi}^2$ (see Eq.~(\ref{eq:ppPhiPhi:scaling})) we restrict our simulation to the case for $N_\phi = 1$. The signal events are further processed for the parton shower and hadronization by {\sc Pythia} v6.4~\cite{Sjostrand:2006za}. We have clustered all particles in the event by {\sc Fastjet 3.1.3}~\cite{Cacciari:2005hq} using the anti-$k_t$ algorithim~\cite{Cacciari:2008gp} with a jet size of $R_{\rm jet} = 0.4$. Only events with at least two jets satisfying $p_T(j) > 20$ GeV and $|\eta(j)| < 4.5$ are selected. 

%%%%%%%%%%%%%%%%%%%%%%%%%%%%
\begin{table}[t]
\centering
\scalebox{0.68}{
\renewcommand{\arraystretch}{1.3}
\begin{tabular}{c|ccccc|ccccc}  
\hline
 & \multicolumn{5}{c|}{$N_{\rm jet}=2$, $|\Delta\phi_{jj}| < 1$, $m_{jj}$ bins} & \multicolumn{5}{c}{$N_{\rm jet}=2$, $1<|\Delta\phi_{jj}| < 2$, $m_{jj}$ bins}
\\
$m_{jj}$ [TeV]  & \quad $0.8-1.0$  & \quad  $1.0-1.5$ & \quad $1.5-2.0$ & \quad $2.0-3.5$ & \quad $> 3.5$ & \quad $0.8-1.0$  & \quad  $1.0-1.5$ & \quad $1.5-2.0$ & \quad $2.0-3.5$  & \quad $> 3.5$
\\
Process & Bin 1 & Bin 2 & Bin 3 & Bin 4 & Bin 5 & Bin 6 & Bin 7 & Bin 8 & Bin 9 & Bin 10
\\
\hline
Total Bkg   & 2040$\pm$44 & 2647$\pm$53  & 884$\pm$28 & 641$\pm$25 & 79$\pm$8 & 1365$\pm$36 & 2728$\pm$52 & 1115$\pm$31 & 842$\pm$28 & 129$\pm$10 \\[2.5pt]
\hline
Data & 2065 & 2639 & 890 & 633 & 76 & 1362 & 2730 & 1132 & 836 & 133 \\[2.5pt]
\hline
\end{tabular}
}
\caption{\small The number of events and associated total systematic uncertainties of total SM backgrounds and observed data in the signal region of $E_T^{\rm miss}+$jets via VBF at $\sqrt{s}$ =13 TeV with an integrated luminosity of 139 fb$^{-1}$. The numbers were taken from Ref.~\cite{ATLAS-CONF-2020-008}.}
\label{tab:ATLAS}
\end{table}
%%%%%%%%%%%%%%%%%%%%%%%%%%%

We impose the following set of kinematic cuts used in the ATLAS analysis~\cite{ATLAS-CONF-2020-008} on our signal events
\begin{equation}\label{app:eq:atlas:cuts}
\begin{split}
& p_T(j_1) > 80\ \text{GeV}~, \quad p_T(j_2) > 50\ \text{GeV}~,
 \\[2.5pt]
& E_T^{\rm miss} > 200\ \text{GeV}~,\quad H_T^{\rm miss} > 180\ \text{GeV}~,\quad m_{jj} > 800\ \text{GeV}~,
 \\[2.5pt]
& \Delta\phi_{jj} < 2.0~,\quad \eta^{j_1}\cdot \eta^{j_2} < 0~, \quad \Delta\eta_{jj} > 3.8~,
\\[2.5pt]
& \text{Centrality: } C_{3,\, 4} < 0.6~, \quad m^{\rm rel}_{3,\, 4} < 0.05~.
\end{split}
\end{equation}
Here  $E_T^{\rm miss}$ is defined as the magnitude of the negative vectorial sum of all reconstructed jets with $p_T(j) > 20$ GeV and $H_T^{\rm miss}$ as the magnitude of the vectorial sum of the transverse momenta of all jets with $p_T > 20$ GeV. The variables $\Delta\phi_{jj}$ and $\Delta\eta_{jj}$ are difference in the azimuthal angle and pseudo-rapidity between two leading jets. $m_{jj}$ is the invariant mass of two leading jets. The centrality $C_{i=3,4}$ and $m^{\rm rel}_{i=3,4}$ of the third and fourth highest jets are defined as 
\begin{equation}
   C_i = \exp \left ( - \frac{4}{(\eta^{j_1} - \eta^{j_2})^2} \left ( \eta^i - \frac{\eta^{j_1}+\eta^{j_2}}{2}\right )^2 \right )~,\quad
   m^{\rm rel}_i = \frac{{\rm min} (m_{j_1,i},m_{j_2,i})}{m_{jj}}~,
\end{equation}
where only the third and fourth jets (if they exist) with $p_T(j) > 25$ GeV are considered and $m_{j_{1},i}$ ($m_{j_{2},i}$) is the invariant mass of the leading jet (the second hardest jet) and $i$-th hardest jet. 

%%%%%%%%%%%%%%%%%%%%%%%
\begin{figure}[htb]
\begin{center}
\includegraphics[width=0.50\textwidth]{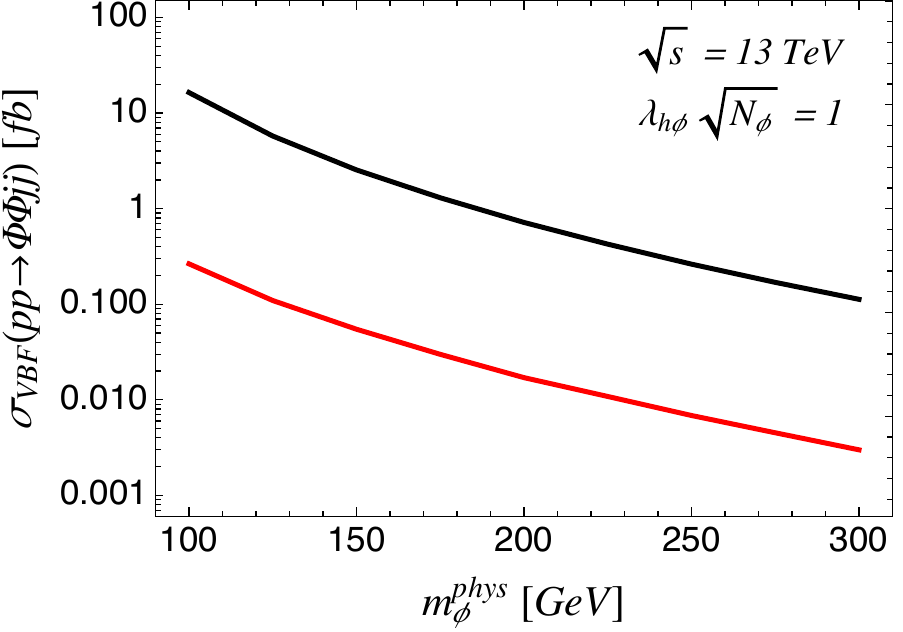}
\caption{\small The inclusive cross sections at the generation level (black) and after imposing kinematic cuts in~\cite{ATLAS-CONF-2020-008} (red) (see Eq.~(\ref{app:eq:atlas:cuts})) as a function of the physical $\Phi$ mass, $m_\phi^{\rm phys}$ for $\lambda_{h\phi} \sqrt{N_\phi} =1$. }
\label{app:fig:xsecS:mass}
\end{center}
\end{figure}
%%%%%%%%%%%%%%%%%%%%%%%

The inclusive cross section after imposing the kinematic cuts in Eq.~(\ref{app:eq:atlas:cuts}) is shown in Fig.~\ref{app:fig:xsecS:mass}. For comparison, we also present in Fig.~\ref{app:fig:xsecS:mass} the inclusive partonic cross section for the singlet $\Phi$ pair production with $\lambda_{h\phi} \sqrt{N_\phi} = 1$. The difference between two curves in Fig.~\ref{app:fig:xsecS:mass} amounts to the signal efficiency. To increase the sensitivity, the events in the ATLAS analysis are binned in $m_{jj}$, $\Delta\phi_{jj}$, and $N_{\rm jet}$ where $N_{\rm jet}$ counts only jets with $p_T(j) > 25$ GeV. The estimates of the total SM backgrounds and the observed events in each bin using 139 fb$^{-1}$ of data in the ATLAS analysis~\cite{ATLAS-CONF-2020-008} are presented in Table~\ref{tab:ATLAS}. While the analysis in~\cite{ATLAS-CONF-2020-008} has the category of $2 < N_{\rm jet} < 5$ with the cut $m_{jj} > 3.5$ TeV, we will not use it as our signal events in the corresponding bin are not significant. Noting that the size of the systematic uncertainty in Table~\ref{tab:ATLAS} is comparable with the statistical uncertainty.

We first derive model-independent limit on the cross section (or the number of events) by recasting each bin of the ATLAS analysis at a time.
Since the number of total backgrounds and data are large enough to ensure the Gaussian distribution due to the central limit theorem, the chi-square from the likelihood $L$ for each bin is given by  
\begin{equation}
 \chi_{\rm bin}^2 \equiv -2 \log L_{\rm bin} = \frac{1}{\sigma_{\rm bin}^2} \Big [ \mathcal{L} \cdot \left ( \sigma_{\rm SM, bin} + \sigma_{\rm NP, bin} \right ) - N_{\rm bin}^{\rm obs} \Big ]^2~,
\end{equation}
where $\mathcal{L}$ denotes the integrated luminosity, $\sigma_{\rm SM, bin}$ is the contribution from the SM, namely total backgrounds to the corresponding bin, and $\sigma_{\rm NP, bin}$ is the cross section from new physics. The variance $\sigma_{\rm bin}$ is taken as the square-root of the squared sum of systematic and statistical uncertainties. The recasted upper limit on $\sigma_{\rm NP}$ (or the number of events with $\mathcal{L}=139$ fb$^{-1}$) at the 95\% CL for each bin is shown in Table~\ref{tab:ATLAS:recast}.

%%%%%%%%%%%%%%%%%%%%%%%%%%%%
\begin{table}[t]
\centering
\scalebox{0.9}{
\renewcommand{\arraystretch}{1.5} 
\begin{tabular}{c|ccccc|ccccc}  
\hline
Process & Bin 1 & Bin 2 & Bin 3 & Bin 4 & Bin 5 & Bin 6 & Bin 7 & Bin 8 & Bin 9 & Bin 10
\\
\hline
$\sigma_{\rm NP, 95\% CL}$ [fb] & 1.07 & 0.984 & 0.619 & 0.444 & 0.147 &  0.706 & 1.05 & 0.765 & 0.525 & 0.242 \\[2.5pt]
$(\mathcal{L}\cdot \sigma_{\rm NP})_{95\% CL}$ & 149 & 137 & 86.0 & 61.7 & 20.4 & 98.1 & 146 & 106 & 73.0 & 33.7 \\[2.5pt]
\hline
\end{tabular}
}
\caption{\small Recasted 95\% CL limits on the cross sections and the numbers of events from new physics contribution to $E_T^{\rm miss}+$jets via VBF at $\sqrt{s}$ =13 TeV with an integrated luminosity of 139 fb$^{-1}$. The limits are derived by considering each bin at a time assuming no correlation between bins.}
\label{tab:ATLAS:recast}
\end{table}

%%%%%%%%%%%%%%%%%%%%%%%
\begin{figure}[thb!]
\begin{center}
\includegraphics[width=0.48\textwidth]{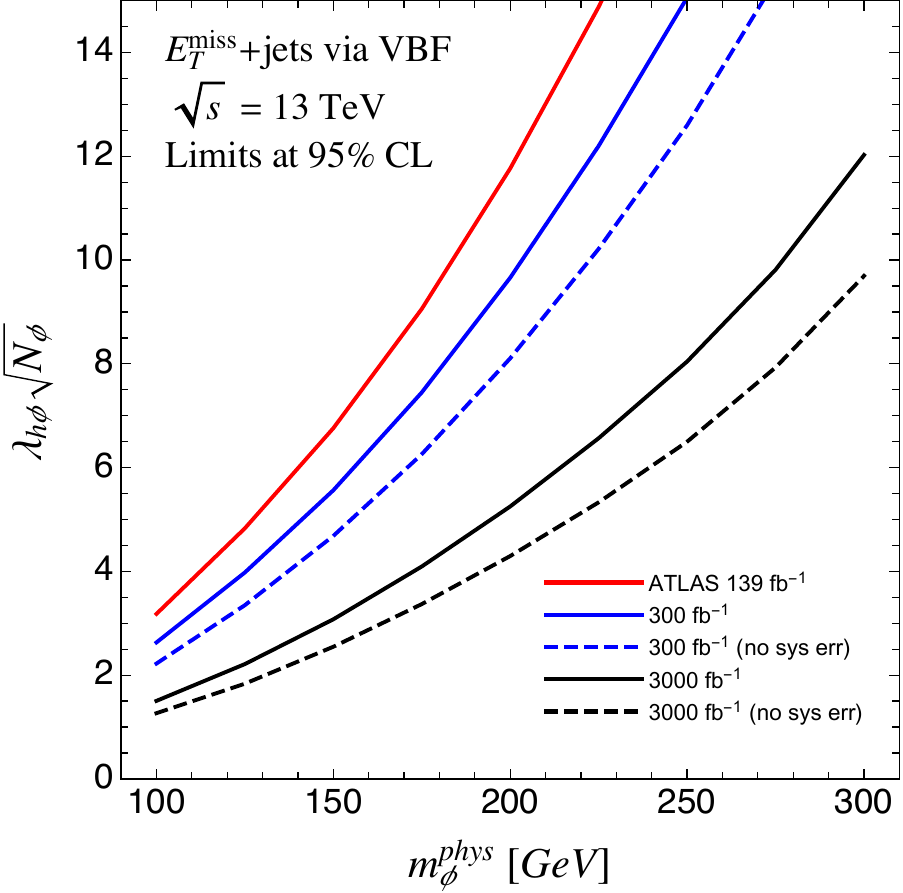}
\caption{\small Recasted limits at the 95\% CL on the portal coupling $\lambda_{h\phi}\sqrt{N_\phi}$ as a function of the physical $\Phi$ mass $m_\phi^{\rm phys}$ based on the ATLAS analysis~\cite{ATLAS-CONF-2020-008} at the 13 TeV LHC with 139 fb$^{-1}$ (red). The projected sensitivities for 300 fb$^{-1}$ (blue) and 3000 fb$^{-1}$ (black) luminosities are obtained based on the ATLAS analysis and assuming both systematic and statistical uncertainties scaling down with the square-root of the luminosity (dashed lines correspond to the cases with only  statistical uncertainties). 
}
\label{app:fig:Etjets:exclusion1}
\end{center}
\end{figure}
%%%%%%%%%%%%%%%%%%%%%%%

In the Higgs portal scenario, the signal events for given parameters $(m^{\rm phys}_\phi,\, \lambda_{h\phi}\sqrt{N_\phi})$ which control the singlet mass and the interaction strength for the production, contributes to all $m_{jj}$ bins. Combining results from all bins will improve the sensitivity on the model parameters. After imposing the kinematic cuts in Eq.~(\ref{app:eq:atlas:cuts}) on the signal events from the singlet production with various masses, we construct the signal cross section as a function of the physical mass in each bin. The total chi-square is given by
\begin{equation}
 \chi^2 \equiv -2 \log L = \sum_{\rm bin}\frac{1}{\sigma_{\rm bin}^2} \Big [ \mathcal{L} \cdot \left ( \sigma_{\rm SM, bin} + N_\phi  \lambda_{h\phi}^2 \cdot \sigma_{\rm singlet, bin}(m^{\rm phys}_\phi ) \right ) - N_{\rm bin}^{\rm obs} \Big ]^2~,
\end{equation}
where $\sigma_{\rm singlet, bin}(m^{\rm phys}_\phi)$ is the cross section as a function of $m_\phi^{\rm phys}$ for $\lambda_{h\phi} \sqrt{N_\phi} = 1$.
The recasted senstivity of the ATLAS analysis~\cite{ATLAS-CONF-2020-008} on the Higgs portal model in the plane $(m^{\rm phys}_\phi,\, \lambda_{h\phi}\sqrt{N_\phi})$ is illustrated in Fig.~\ref{app:fig:Etjets:exclusion1}. For the projected sensitivities at the LHC with more luminosities, we make simple projections for integrated luminosities of 300 fb$^{-1}$ and 3000 fb$^{-1}$ assuming that both systematic and statistical uncertainties scale as $1/\sqrt{\mathcal{L}}$. To understand the impact of the systematic uncertainties on the sensitivities, we also estimate the limits for the cases with only statistical uncertainties. The resulting limits at the 95\% CL are shown in Fig.~\ref{app:fig:Etjets:exclusion1}.

%
%\bibliographystyle{JHEP}
%\bibliography{references}
%%%%%%%%%%%%%%%%%
% References
%%%%%%%%%%%%%%%%%
%{\small
%\bibliography{lit}{}

\begin{thebibliography}{10}

\bibitem{ATLAS:2012ulh}
{\scshape ATLAS} collaboration, G.~Aad et~al., \emph{{Observation of a new
  particle in the search for the Standard Model Higgs boson with the ATLAS
  detector at the LHC}},
  \href{https://doi.org/10.1016/j.physletb.2012.08.020}{\emph{Phys. Lett. B}
  {\bfseries 716} (2012) 1} [\href{https://arxiv.org/abs/1207.7214}{{\ttfamily
  1207.7214}}].

\bibitem{CMS:2012rpq}
{\scshape CMS} collaboration, S.~Chatrchyan et~al., \emph{{Observation of a New
  Boson at a Mass of 125 GeV with the CMS Experiment at the LHC}},
  \href{https://doi.org/10.1016/j.physletb.2012.08.021}{\emph{Phys. Lett. B}
  {\bfseries 716} (2012) 30} [\href{https://arxiv.org/abs/1207.7235}{{\ttfamily
  1207.7235}}].

\bibitem{Bousso:2000xa}
R.~Bousso and J.~Polchinski, \emph{{Quantization of four form fluxes and
  dynamical neutralization of the cosmological constant}},
  \href{https://doi.org/10.1088/1126-6708/2000/06/006}{\emph{JHEP} {\bfseries
  06} (2000) 006} [\href{https://arxiv.org/abs/hep-th/0004134}{{\ttfamily
  hep-th/0004134}}].

\bibitem{Kachru:2003aw}
S.~Kachru, R.~Kallosh, A.~D. Linde and S.~P. Trivedi, \emph{{De Sitter vacua in
  string theory}},
  \href{https://doi.org/10.1103/PhysRevD.68.046005}{\emph{Phys. Rev. D}
  {\bfseries 68} (2003) 046005}
  [\href{https://arxiv.org/abs/hep-th/0301240}{{\ttfamily hep-th/0301240}}].

\bibitem{Susskind:2003kw}
L.~Susskind, \emph{{The Anthropic landscape of string theory}},
  \href{https://arxiv.org/abs/hep-th/0302219}{{\ttfamily hep-th/0302219}}.

\bibitem{Douglas:2003um}
M.~R. Douglas, \emph{{The Statistics of string / M theory vacua}},
  \href{https://doi.org/10.1088/1126-6708/2003/05/046}{\emph{JHEP} {\bfseries
  05} (2003) 046} [\href{https://arxiv.org/abs/hep-th/0303194}{{\ttfamily
  hep-th/0303194}}].

\bibitem{Sher:1988mj}
M.~Sher, \emph{{Electroweak Higgs Potentials and Vacuum Stability}},
  \href{https://doi.org/10.1016/0370-1573(89)90061-6}{\emph{Phys. Rept.}
  {\bfseries 179} (1989) 273}.

\bibitem{Degrassi:2012ry}
G.~Degrassi, S.~Di~Vita, J.~Elias-Miro, J.~R. Espinosa, G.~F. Giudice,
  G.~Isidori et~al., \emph{{Higgs mass and vacuum stability in the Standard
  Model at NNLO}}, \href{https://doi.org/10.1007/JHEP08(2012)098}{\emph{JHEP}
  {\bfseries 08} (2012) 098} [\href{https://arxiv.org/abs/1205.6497}{{\ttfamily
  1205.6497}}].

\bibitem{Quigg:2009xr}
C.~Quigg and R.~Shrock, \emph{{Gedanken Worlds without Higgs: QCD-Induced
  Electroweak Symmetry Breaking}},
  \href{https://doi.org/10.1103/PhysRevD.79.096002}{\emph{Phys. Rev. D}
  {\bfseries 79} (2009) 096002}
  [\href{https://arxiv.org/abs/0901.3958}{{\ttfamily 0901.3958}}].

\bibitem{Weinberg:1974hy}
S.~Weinberg, \emph{{Gauge and Global Symmetries at High Temperature}},
  \href{https://doi.org/10.1103/PhysRevD.9.3357}{\emph{Phys. Rev. D} {\bfseries
  9} (1974) 3357}.

\bibitem{Cohen:1993nk}
A.~G. Cohen, D.~B. Kaplan and A.~E. Nelson, \emph{{Progress in electroweak
  baryogenesis}},
  \href{https://doi.org/10.1146/annurev.ns.43.120193.000331}{\emph{Ann. Rev.
  Nucl. Part. Sci.} {\bfseries 43} (1993) 27}
  [\href{https://arxiv.org/abs/hep-ph/9302210}{{\ttfamily hep-ph/9302210}}].

\bibitem{Cline:2006ts}
J.~M. Cline, \emph{{Baryogenesis}},  in \emph{{Les Houches Summer School -
  Session 86: Particle Physics and Cosmology: The Fabric of Spacetime}}, 9,
  2006, \href{https://arxiv.org/abs/hep-ph/0609145}{{\ttfamily
  hep-ph/0609145}}.

\bibitem{Meade:2018saz}
P.~Meade and H.~Ramani, \emph{{Unrestored Electroweak Symmetry}},
  \href{https://doi.org/10.1103/PhysRevLett.122.041802}{\emph{Phys. Rev. Lett.}
  {\bfseries 122} (2019) 041802}
  [\href{https://arxiv.org/abs/1807.07578}{{\ttfamily 1807.07578}}].

\bibitem{Baldes:2018nel}
I.~Baldes and G.~Servant, \emph{{High scale electroweak phase transition:
  baryogenesis and symmetry non-restoration}},
  \href{https://doi.org/10.1007/JHEP10(2018)053}{\emph{JHEP} {\bfseries 10}
  (2018) 053} [\href{https://arxiv.org/abs/1807.08770}{{\ttfamily
  1807.08770}}].

\bibitem{Glioti:2018roy}
A.~Glioti, R.~Rattazzi and L.~Vecchi, \emph{{Electroweak Baryogenesis above the
  Electroweak Scale}},
  \href{https://doi.org/10.1007/JHEP04(2019)027}{\emph{JHEP} {\bfseries 04}
  (2019) 027} [\href{https://arxiv.org/abs/1811.11740}{{\ttfamily
  1811.11740}}].

\bibitem{Matsedonskyi:2020mlz}
O.~Matsedonskyi and G.~Servant, \emph{{High-Temperature Electroweak Symmetry
  Non-Restoration from New Fermions and Implications for Baryogenesis}},
  \href{https://doi.org/10.1007/JHEP09(2020)012}{\emph{JHEP} {\bfseries 09}
  (2020) 012} [\href{https://arxiv.org/abs/2002.05174}{{\ttfamily
  2002.05174}}].

\bibitem{Cao:2021yau}
Q.-H. Cao, K.~Hashino, X.-X. Li, Z.~Ren and J.-H. Yu, \emph{{Electroweak phase
  transition triggered by fermion sector}},
  \href{https://arxiv.org/abs/2103.05688}{{\ttfamily 2103.05688}}.

\bibitem{Langacker:1980kd}
P.~Langacker and S.-Y. Pi, \emph{{Magnetic Monopoles in Grand Unified
  Theories}}, \href{https://doi.org/10.1103/PhysRevLett.45.1}{\emph{Phys. Rev.
  Lett.} {\bfseries 45} (1980) 1}.

\bibitem{Salomonson:1984rh}
P.~Salomonson, B.~S. Skagerstam and A.~Stern, \emph{{On the Primordial Monopole
  Problem in Grand Unified Theories}},
  \href{https://doi.org/10.1016/0370-2693(85)90843-3}{\emph{Phys. Lett. B}
  {\bfseries 151} (1985) 243}.

\bibitem{Dvali:1995cj}
G.~R. Dvali, A.~Melfo and G.~Senjanovic, \emph{{Is There a monopole problem?}},
  \href{https://doi.org/10.1103/PhysRevLett.75.4559}{\emph{Phys. Rev. Lett.}
  {\bfseries 75} (1995) 4559}
  [\href{https://arxiv.org/abs/hep-ph/9507230}{{\ttfamily hep-ph/9507230}}].

\bibitem{Dvali:1995cc}
G.~R. Dvali and G.~Senjanovic, \emph{{Is there a domain wall problem?}},
  \href{https://doi.org/10.1103/PhysRevLett.74.5178}{\emph{Phys. Rev. Lett.}
  {\bfseries 74} (1995) 5178}
  [\href{https://arxiv.org/abs/hep-ph/9501387}{{\ttfamily hep-ph/9501387}}].

\bibitem{Chai:2020zgq}
N.~Chai, S.~Chaudhuri, C.~Choi, Z.~Komargodski, E.~Rabinovici and M.~Smolkin,
  \emph{{Thermal Order in Conformal Theories}},
  \href{https://doi.org/10.1103/PhysRevD.102.065014}{\emph{Phys. Rev. D}
  {\bfseries 102} (2020) 065014}
  [\href{https://arxiv.org/abs/2005.03676}{{\ttfamily 2005.03676}}].

\bibitem{Mohapatra:1979qt}
R.~N. Mohapatra and G.~Senjanovic, \emph{{Soft CP Violation at High
  Temperature}}, \href{https://doi.org/10.1103/PhysRevLett.42.1651}{\emph{Phys.
  Rev. Lett.} {\bfseries 42} (1979) 1651}.

\bibitem{Mohapatra:1979vr}
R.~N. Mohapatra and G.~Senjanovic, \emph{{Broken Symmetries at High
  Temperature}}, \href{https://doi.org/10.1103/PhysRevD.20.3390}{\emph{Phys.
  Rev. D} {\bfseries 20} (1979) 3390}.

\bibitem{Bodeker:2017cim}
D.~Bodeker and G.~D. Moore, \emph{{Electroweak Bubble Wall Speed Limit}},
  \href{https://doi.org/10.1088/1475-7516/2017/05/025}{\emph{JCAP} {\bfseries
  05} (2017) 025} [\href{https://arxiv.org/abs/1703.08215}{{\ttfamily
  1703.08215}}].

\bibitem{Hoeche:2020rsg}
S.~H\"oche, J.~Kozaczuk, A.~J. Long, J.~Turner and Y.~Wang, \emph{{Towards an
  all-orders calculation of the electroweak bubble wall velocity}},
  \href{https://doi.org/10.1088/1475-7516/2021/03/009}{\emph{JCAP} {\bfseries
  03} (2021) 009} [\href{https://arxiv.org/abs/2007.10343}{{\ttfamily
  2007.10343}}].

\bibitem{Vanvlasselaer:2020niz}
A.~Azatov and M.~Vanvlasselaer, \emph{{Bubble wall velocity: heavy physics
  effects}}, \href{https://doi.org/10.1088/1475-7516/2021/01/058}{\emph{JCAP}
  {\bfseries 01} (2021) 058}
  [\href{https://arxiv.org/abs/2010.02590}{{\ttfamily 2010.02590}}].

\bibitem{Dolan:1973qd}
L.~Dolan and R.~Jackiw, \emph{{Symmetry Behavior at Finite Temperature}},
  \href{https://doi.org/10.1103/PhysRevD.9.3320}{\emph{Phys. Rev. D} {\bfseries
  9} (1974) 3320}.

\bibitem{Kraemmer:2003gd}
U.~Kraemmer and A.~Rebhan, \emph{{Advances in perturbative thermal field
  theory}}, \href{https://doi.org/10.1088/0034-4885/67/3/R05}{\emph{Rept. Prog.
  Phys.} {\bfseries 67} (2004) 351}
  [\href{https://arxiv.org/abs/hep-ph/0310337}{{\ttfamily hep-ph/0310337}}].

\bibitem{Delaunay:2007wb}
C.~Delaunay, C.~Grojean and J.~D. Wells, \emph{{Dynamics of Non-renormalizable
  Electroweak Symmetry Breaking}},
  \href{https://doi.org/10.1088/1126-6708/2008/04/029}{\emph{JHEP} {\bfseries
  04} (2008) 029} [\href{https://arxiv.org/abs/0711.2511}{{\ttfamily
  0711.2511}}].

\bibitem{Jain:2017sqm}
B.~Jain, S.~J. Lee and M.~Son, \emph{{Validity of the effective potential and
  the precision of Higgs field self-couplings}},
  \href{https://doi.org/10.1103/PhysRevD.98.075002}{\emph{Phys. Rev. D}
  {\bfseries 98} (2018) 075002}
  [\href{https://arxiv.org/abs/1709.03232}{{\ttfamily 1709.03232}}].

\bibitem{Coleman:1977py}
S.~R. Coleman, \emph{{The Fate of the False Vacuum. 1. Semiclassical Theory}},
  \href{https://doi.org/10.1103/PhysRevD.16.1248}{\emph{Phys. Rev. D}
  {\bfseries 15} (1977) 2929}.

\bibitem{Linde:1980tt}
A.~D. Linde, \emph{{Fate of the False Vacuum at Finite Temperature: Theory and
  Applications}},
  \href{https://doi.org/10.1016/0370-2693(81)90281-1}{\emph{Phys. Lett. B}
  {\bfseries 100} (1981) 37}.

\bibitem{Mathematica}
W.~R. Inc., ``Mathematica, {V}ersion 12.2.''

\bibitem{Guada:2020xnz}
V.~Guada, M.~Nemev\v{s}ek and M.~Pintar, \emph{{FindBounce: Package for
  multi-field bounce actions}},
  \href{https://doi.org/10.1016/j.cpc.2020.107480}{\emph{Comput. Phys. Commun.}
  {\bfseries 256} (2020) 107480}
  [\href{https://arxiv.org/abs/2002.00881}{{\ttfamily 2002.00881}}].

\bibitem{Espinosa:2015qea}
J.~R. Espinosa, G.~F. Giudice, E.~Morgante, A.~Riotto, L.~Senatore, A.~Strumia
  et~al., \emph{{The cosmological Higgstory of the vacuum instability}},
  \href{https://doi.org/10.1007/JHEP09(2015)174}{\emph{JHEP} {\bfseries 09}
  (2015) 174} [\href{https://arxiv.org/abs/1505.04825}{{\ttfamily
  1505.04825}}].

\bibitem{Coleman:1980aw}
S.~R. Coleman and F.~De~Luccia, \emph{{Gravitational Effects on and of Vacuum
  Decay}}, \href{https://doi.org/10.1103/PhysRevD.21.3305}{\emph{Phys. Rev. D}
  {\bfseries 21} (1980) 3305}.

\bibitem{Guth:1981uk}
A.~H. Guth and E.~J. Weinberg, \emph{{Cosmological Consequences of a First
  Order Phase Transition in the SU(5) Grand Unified Model}},
  \href{https://doi.org/10.1103/PhysRevD.23.876}{\emph{Phys. Rev. D} {\bfseries
  23} (1981) 876}.

\bibitem{Gleiser:1991rf}
M.~Gleiser, E.~W. Kolb and R.~Watkins, \emph{{Phase transitions with
  subcritical bubbles}},
  \href{https://doi.org/10.1016/0550-3213(91)90592-L}{\emph{Nucl. Phys. B}
  {\bfseries 364} (1991) 411}.

\bibitem{Gleiser:1995er}
M.~Gleiser, A.~F. Heckler and E.~W. Kolb, \emph{{Modeling thermal fluctuations:
  Phase mixing and percolation}},
  \href{https://doi.org/10.1016/S0370-2693(97)00621-7}{\emph{Phys. Lett. B}
  {\bfseries 405} (1997) 121}
  [\href{https://arxiv.org/abs/cond-mat/9512032}{{\ttfamily
  cond-mat/9512032}}].

\bibitem{Friedberg:1976me}
R.~Friedberg, T.~Lee and A.~Sirlin, \emph{{A Class of Scalar-Field Soliton
  Solutions in Three Space Dimensions}},
  \href{https://doi.org/10.1103/PhysRevD.13.2739}{\emph{Phys. Rev. D}
  {\bfseries 13} (1976) 2739}.

\bibitem{Coleman:1985ki}
S.~R. Coleman, \emph{{Q Balls}},
  \href{https://doi.org/10.1016/0550-3213(86)90520-1}{\emph{Nucl. Phys. B}
  {\bfseries 262} (1985) 263}.

\bibitem{Ponton:2019hux}
E.~Pont\'on, Y.~Bai and B.~Jain, \emph{{Electroweak Symmetric Dark Matter
  Balls}}, \href{https://doi.org/10.1007/s13130-019-11194-5}{\emph{JHEP}
  {\bfseries 09} (2019) 011}
  [\href{https://arxiv.org/abs/1906.10739}{{\ttfamily 1906.10739}}].

\bibitem{Sirunyan:2018ayu}
{\scshape CMS} collaboration, A.~M. Sirunyan et~al., \emph{{Combination of
  searches for Higgs boson pair production in proton-proton collisions at
  $\sqrt{s} = $ 13 TeV}},
  \href{https://doi.org/10.1103/PhysRevLett.122.121803}{\emph{Phys. Rev. Lett.}
  {\bfseries 122} (2019) 121803}
  [\href{https://arxiv.org/abs/1811.09689}{{\ttfamily 1811.09689}}].

\bibitem{Aad:2019uzh}
{\scshape ATLAS} collaboration, G.~Aad et~al., \emph{{Combination of searches
  for Higgs boson pairs in $pp$ collisions at $\sqrt{s} = $13 TeV with the
  ATLAS detector}},
  \href{https://doi.org/10.1016/j.physletb.2019.135103}{\emph{Phys. Lett. B}
  {\bfseries 800} (2020) 135103}
  [\href{https://arxiv.org/abs/1906.02025}{{\ttfamily 1906.02025}}].

\bibitem{DiMicco:2019ngk}
J.~Alison et~al., \emph{{Higgs boson potential at colliders: Status and
  perspectives}}, \href{https://doi.org/10.1016/j.revip.2020.100045}{\emph{Rev.
  Phys.} {\bfseries 5} (2020) 100045}
  [\href{https://arxiv.org/abs/1910.00012}{{\ttfamily 1910.00012}}].

\bibitem{ATL-PHYS-PUB-2019-009}
{\scshape ATLAS} collaboration, \emph{{Constraint of the Higgs boson
  self-coupling from Higgs boson differential production and decay
  measurements}},  Tech. Rep. ATL-PHYS-PUB-2019-009, CERN, Geneva, Mar, 2019.

\bibitem{CMS-PAS-HIG-19-005}
{\scshape CMS} collaboration, \emph{{Combined Higgs boson production and decay
  measurements with up to 137 fb$^{-1}$ of proton-proton collision data at
  sqrts = 13 TeV}},  Tech. Rep. CMS-PAS-HIG-19-005, CERN, Geneva, 2020.

\bibitem{Cepeda:2019klc}
M.~Cepeda et~al., \emph{{Report from Working Group 2}: {Higgs Physics at the
  HL-LHC and HE-LHC}},
  \href{https://doi.org/10.23731/CYRM-2019-007.221}{\emph{CERN Yellow Rep.
  Monogr.} {\bfseries 7} (2019) 221}
  [\href{https://arxiv.org/abs/1902.00134}{{\ttfamily 1902.00134}}].

\bibitem{DiVita:2017vrr}
S.~Di~Vita, G.~Durieux, C.~Grojean, J.~Gu, Z.~Liu, G.~Panico et~al., \emph{{A
  global view on the Higgs self-coupling at lepton colliders}},
  \href{https://doi.org/10.1007/JHEP02(2018)178}{\emph{JHEP} {\bfseries 02}
  (2018) 178} [\href{https://arxiv.org/abs/1711.03978}{{\ttfamily
  1711.03978}}].

\bibitem{ATLAS-CONF-2020-008}
{\scshape ATLAS} collaboration, \emph{{Search for invisible Higgs boson decays
  with vector boson fusion signatures with the ATLAS detector using an
  integrated luminosity of 139 fb$^{-1}$}},  Tech. Rep. ATLAS-CONF-2020-008,
  CERN, Geneva, Apr, 2020.

\bibitem{Papaefstathiou:2015paa}
A.~Papaefstathiou and K.~Sakurai, \emph{{Triple Higgs boson production at a 100
  TeV proton-proton collider}},
  \href{https://doi.org/10.1007/JHEP02(2016)006}{\emph{JHEP} {\bfseries 02}
  (2016) 006} [\href{https://arxiv.org/abs/1508.06524}{{\ttfamily
  1508.06524}}].

\bibitem{Chen:2015gva}
C.-Y. Chen, Q.-S. Yan, X.~Zhao, Y.-M. Zhong and Z.~Zhao, \emph{{Probing
  triple-Higgs productions via 4b2\ensuremath{\gamma} decay channel at a 100
  TeV hadron collider}},
  \href{https://doi.org/10.1103/PhysRevD.93.013007}{\emph{Phys. Rev. D}
  {\bfseries 93} (2016) 013007}
  [\href{https://arxiv.org/abs/1510.04013}{{\ttfamily 1510.04013}}].

\bibitem{Fuks:2017zkg}
B.~Fuks, J.~H. Kim and S.~J. Lee, \emph{{Scrutinizing the Higgs quartic
  coupling at a future 100 TeV proton\textendash{}proton collider with taus and
  b-jets}}, \href{https://doi.org/10.1016/j.physletb.2017.05.075}{\emph{Phys.
  Lett. B} {\bfseries 771} (2017) 354}
  [\href{https://arxiv.org/abs/1704.04298}{{\ttfamily 1704.04298}}].

\bibitem{Kilian:2017nio}
W.~Kilian, S.~Sun, Q.-S. Yan, X.~Zhao and Z.~Zhao, \emph{{New Physics in
  multi-Higgs boson final states}},
  \href{https://doi.org/10.1007/JHEP06(2017)145}{\emph{JHEP} {\bfseries 06}
  (2017) 145} [\href{https://arxiv.org/abs/1702.03554}{{\ttfamily
  1702.03554}}].

\bibitem{Papaefstathiou:2019ofh}
A.~Papaefstathiou, G.~Tetlalmatzi-Xolocotzi and M.~Zaro, \emph{{Triple Higgs
  boson production to six $b$-jets at a 100 TeV proton collider}},
  \href{https://doi.org/10.1140/epjc/s10052-019-7457-1}{\emph{Eur. Phys. J. C}
  {\bfseries 79} (2019) 947}
  [\href{https://arxiv.org/abs/1909.09166}{{\ttfamily 1909.09166}}].

\bibitem{Maltoni:2018ttu}
F.~Maltoni, D.~Pagani and X.~Zhao, \emph{{Constraining the Higgs 
 self-couplings at e$^{+}$e$^{−}$ colliders}},
 \href{https://doi.org/10.1007/JHEP07(2018)087}{\emph{JHEP}
 {\bfseries 07} (2018) 087}
 [\href{https://arxiv.org/abs/1802.07616}{{\ttfamily 1802.07616}}].

\bibitem{Liu:2018peg}
T.~Liu, K.~F.~Lyu, J.~Ren and H.~X.~Zhu, \emph{{Probing the quartic Higgs 
 boson self-interaction}},
 \href{https://doi.org/10.1103/PhysRevD.98.093004}{\emph{Phys. Rev. D}
 {\bfseries 98} (2018) 093004}
 [\href{https://arxiv.org/abs/1803.04359}{{\ttfamily 1803.04359}}].

\bibitem{Borowka:2018pxx}
S.~Borowka, C.~Duhr, F.~Maltoni, D.~Pagani, A.~Shivaji and X.~Zhao, 
\emph{{Probing the scalar potential via double Higgs boson production at 
 hadron colliders}},
 \href{https://doi.org/10.1007/JHEP04(2019)016}{\emph{JHEP}
 {\bfseries 04} (2019) 016}
 [\href{https://arxiv.org/abs/1811.12366}{{\ttfamily 1811.12366}}].

\bibitem{Bizon:2018syu}
W.~Bizo\'n, U.~Haisch and L.~Rottoli, \emph{{Constraints on the quartic Higgs 
 self-coupling from double-Higgs production at future hadron colliders}},
 \href{https://doi.org/10.1007/JHEP10(2019)267}{\emph{JHEP} 
 {\bfseries 10} (2019) 267}
 [\href{https://arxiv.org/abs/1810.04665}{{\ttfamily 1810.04665}}].

\bibitem{Craig:2014lda}
N.~Craig, H.~K. Lou, M.~McCullough and A.~Thalapillil, \emph{{The Higgs Portal
  Above Threshold}}, \href{https://doi.org/10.1007/JHEP02(2016)127}{\emph{JHEP}
  {\bfseries 02} (2016) 127} [\href{https://arxiv.org/abs/1412.0258}{{\ttfamily
  1412.0258}}].

\bibitem{Ruhdorfer:2019utl}
M.~Ruhdorfer, E.~Salvioni and A.~Weiler, \emph{{A Global View of the Off-Shell
  Higgs Portal}},
  \href{https://doi.org/10.21468/SciPostPhys.8.2.027}{\emph{SciPost Phys.}
  {\bfseries 8} (2020) 027} [\href{https://arxiv.org/abs/1910.04170}{{\ttfamily
  1910.04170}}].

\bibitem{Baer:2013cma}
\emph{{The International Linear Collider Technical Design Report - Volume 2:
  Physics}},  \href{https://arxiv.org/abs/1306.6352}{{\ttfamily 1306.6352}}.

\bibitem{CEPCStudyGroup:2018ghi}
{\scshape CEPC Study Group} collaboration, M.~Dong et~al., \emph{{CEPC
  Conceptual Design Report: Volume 2 - Physics and Detector}},
  \href{https://arxiv.org/abs/1811.10545}{{\ttfamily 1811.10545}}.

\bibitem{Abada:2019zxq}
{\scshape FCC} collaboration, A.~Abada et~al., \emph{{FCC-ee: The Lepton
  Collider}: {Future Circular Collider Conceptual Design Report Volume 2}},
  \href{https://doi.org/10.1140/epjst/e2019-900045-4}{\emph{Eur. Phys. J. ST}
  {\bfseries 228} (2019) 261}.

\bibitem{Barger:2007im}
V.~Barger, P.~Langacker, M.~McCaskey, M.~J. Ramsey-Musolf and G.~Shaughnessy,
  \emph{{LHC Phenomenology of an Extended Standard Model with a Real Scalar
  Singlet}}, \href{https://doi.org/10.1103/PhysRevD.77.035005}{\emph{Phys. Rev.
  D} {\bfseries 77} (2008) 035005}
  [\href{https://arxiv.org/abs/0706.4311}{{\ttfamily 0706.4311}}].

\bibitem{Alloul:2013bka}
A.~Alloul, N.~D. Christensen, C.~Degrande, C.~Duhr and B.~Fuks,
  \emph{{FeynRules 2.0 - A complete toolbox for tree-level phenomenology}},
  \href{https://doi.org/10.1016/j.cpc.2014.04.012}{\emph{Comput. Phys. Commun.}
  {\bfseries 185} (2014) 2250}
  [\href{https://arxiv.org/abs/1310.1921}{{\ttfamily 1310.1921}}].

\bibitem{Alwall:2014hca}
J.~Alwall, R.~Frederix, S.~Frixione, V.~Hirschi, F.~Maltoni, O.~Mattelaer
  et~al., \emph{{The automated computation of tree-level and next-to-leading
  order differential cross sections, and their matching to parton shower
  simulations}}, \href{https://doi.org/10.1007/JHEP07(2014)079}{\emph{JHEP}
  {\bfseries 07} (2014) 079} [\href{https://arxiv.org/abs/1405.0301}{{\ttfamily
  1405.0301}}].

\bibitem{Sjostrand:2006za}
T.~Sjostrand, S.~Mrenna and P.~Z. Skands, \emph{{PYTHIA 6.4 Physics and
  Manual}}, \href{https://doi.org/10.1088/1126-6708/2006/05/026}{\emph{JHEP}
  {\bfseries 05} (2006) 026}
  [\href{https://arxiv.org/abs/hep-ph/0603175}{{\ttfamily hep-ph/0603175}}].

\bibitem{Cacciari:2005hq}
M.~Cacciari and G.~P. Salam, \emph{{Dispelling the $N^{3}$ myth for the $k_t$
  jet-finder}},
  \href{https://doi.org/10.1016/j.physletb.2006.08.037}{\emph{Phys. Lett. B}
  {\bfseries 641} (2006) 57}
  [\href{https://arxiv.org/abs/hep-ph/0512210}{{\ttfamily hep-ph/0512210}}].

\bibitem{Cacciari:2008gp}
M.~Cacciari, G.~P. Salam and G.~Soyez, \emph{{The anti-$k_t$ jet clustering
  algorithm}}, \href{https://doi.org/10.1088/1126-6708/2008/04/063}{\emph{JHEP}
  {\bfseries 04} (2008) 063} [\href{https://arxiv.org/abs/0802.1189}{{\ttfamily
  0802.1189}}].

\end{thebibliography}
%\bibliographystyle{JHEP}}

\providecommand{\href}[2]{#2}\begingroup\raggedright\endgroup

\end{document}